\documentclass[pre,twocolumn,a4paper,amsmath,floatfix, nofootinbib]{revtex4-1}
\usepackage[utf8]{inputenc}
\usepackage{graphicx}
\usepackage{amsmath}
\usepackage{numprint}

\usepackage{color}
\newcommand{\bigO}[1]{\mathcal{O}(#1)}
\newcommand{\CELL}{{\bf CellList}}
\newcommand{\MULTI}{{\bf MultiEvent}}
\newcommand{\SINGLE}{{\bf SingleEvent}}

\begin{document}
\title{Efficient event-driven simulations of hard spheres}
\author{Frank Smallenburg}
\affiliation{Universit\'e Paris-Saclay, CNRS, Laboratoire de Physique des Solides, 91405 Orsay, France}

\begin{abstract}
Hard spheres are arguably one of the most fundamental model systems in soft matter physics, and hence a common topic of simulation studies. Event-driven simulation methods provide an efficient method for studying the phase behavior and dynamics of hard spheres under a wide range of different conditions. Here, we examine the impact of several optimization strategies for speeding up event-driven molecular dynamics of hard spheres, and present a light-weight simulation code that outperforms existing simulation codes over a large range of system sizes and packing fractions. The presented differences in simulation speed, typically a factor of five to ten, save significantly on both CPU time and energy consumption, and may be a crucial factor for studying slow processes such as crystal nucleation and glassy dynamics.
\end{abstract}

\maketitle

\section{Introduction}

Hard spheres are perhaps the most commonly studied model system in soft matter physics. In addition to being excellent model systems for certain colloidal suspensions \cite{pusey1986phase,  kegel2000direct, royall2013search}, their simplicity makes them a fundamental model study for many-body systems at any scale. As a result, simulations of hard spheres have been instrumental in shedding light on many aspects of statistical physics, including phase behavior in bulk\cite{frenkel1984new,biben1991phase, dijkstra1998phase, dijkstra1998nonadditive, dijkstra1999phase, filion2011self, bommineni2019complex, bommineni2020spontaneous} and in confinement \cite{schmidt1997phase, fortini2006phase, gordillo2006freezing, ouguz2012packing, de2015entropy,jung2020confinement, wang2021binary}, crystal nucleation\cite{auer2001prediction,punnathanam2006crystal,  filion2010crystal, ni2011crystal, berryman2016early, richard2018crystallization, wood2018coupling, espinosa2019heterogeneous}, and glassy materials \cite{rintoul1996computer, speedy1998hard, leocmach2012roles, sanz2014avalanches, urbani2017shear, berthier2017configurational, lazaro2019glassy, marin2020tetrahedrality, boattini2021averaging, boattini2020autonomously}. 

Due to their popularity as a model system, it should come as no surprise that significant efforts have been made to design efficient hard-sphere simulations, typically based on either Monte Carlo (MC) methods or event-driven molecular dynamics (EDMD). Monte Carlo methods have been applied to hard-sphere systems since the 1950s \cite{rosenbluth1954further, wood1957preliminary} and continue to be in active use. For a hard-sphere system in the canonical ensemble, the basic MC algorithm consists of simple single-particle displacement moves, which are accepted as long as no overlaps are created \cite{bookfrenkel}. This approach can readily be adapted to a variety of thermodynamic ensembles, or to incorporate biasing potentials or external fields, and scales well to large systems. Moreover, intelligently chosen MC trial moves, such as cluster moves\cite{dress1995cluster, almarza2009cluster, ashton2011depletion} or swap moves \cite{berthier2016equilibrium}, can lead to significant speedups in sampling speed. In particular, event-chain algorithms, which move chains of particles collectively in a rejection-free move, are highly efficient for hard-sphere systems \cite{bernard2009event, klement2019efficient}.  Depending on the goal of the simulations, one potential downside of MC simulations is the fact that, unless special care is taken\cite{sanz2010dynamic, cuetos2015equivalence}, the dynamics do not typically correspond well to that of a real-world system. 

Event-driven molecular dynamics (EDMD) simulations are a second widely used method for simulating hard-sphere systems, with a similarly long history \cite{alder1957phase}. These simulations solve for the motion of the particles using Newton's equations of motion, effectively modeling the movement of hard spheres in a vacuum\footnote{It should be noted that variations on EDMD that simulate Brownian\cite{scala2007event} or Langevin\cite{scala2012event} dynamics have been developed as well.}. Following Newton's third law, hard spheres move in simple straight lines until they collide with a neighboring particle. EDMD simulations rely on predicting future collision events between particles, and then resolving each of these collision events in order. In comparison to standard MC simulations, hard-sphere EDMD simulations are extremely efficient, rivaling e.g. event-chain methods \cite{klement2019efficient}, with the added benefit of reproducing physically realistic dynamics.  This is ideal for the study of glassy dynamics or spontaneous nucleation, where observing the dynamics of interest requires simulation over long time scales.

Here, we present a light-weight and fast hard-sphere EDMD code, which combines an efficient event-queueing system \cite{paul2007complexity}, neighbor lists \cite{donev2005neighbor}, and ``lazy'' event evaluation \cite{lubachevsky1991simulate}. We benchmark the simulation code on a range of monodisperse and bidisperse hard-sphere systems, and find significant performance gains when comparing the results with other commonly used simulation codes \cite{bannerman2011dynamo, rapaport2004art}. 

In the remainder of this article, we discuss some of the more detailed aspects of the efficient implementation of EDMD simulations (Sec. \ref{sec:implementation}), show benchmarks of three variations of the presented simulation code (Sec. \ref{sec:benchmarks}), and discuss the results (Sec. \ref{sec:discuss}).

\section{Implementation}
\label{sec:implementation}

In this section, we outline the implementation of an efficient event-driven hard-sphere simulation algorithm. After briefly outlining the typical setup of a hard-sphere EDMD, we focus on three implementation choices: the event calendar, the neighbor selection, and the choice of how many events to schedule for a given particle.  

For the simulations described in this article, we always consider periodic systems in an orthogonal simulation box. 

\subsection{Basic outline}

The basic premise of a hard-sphere EDMD is straightforward. Given a configuration of $N$ particles with positions $\{\mathbf{r}^N\}$ and velocities $\{\mathbf{v}^N\}$, the algorithm determines which pair of particles will be the first to undergo a collision, and when this collision will take place. Once the event is determined, the simulation time $t$ jumps forward to the collision time, particle positions and velocities are updated to reflect the impact of the collision, and the simulation proceeds to predict the next collision. 

Since we are simulating purely hard particles moving in a vacuum, we know that in between collision events the particles experience no forces, and hence move in straight lines. As a result, for hard spheres, predicting the collision time for a pair of particles can be done analytically based on the initial distance vector $\mathbf{r}_{ij} = \mathbf{r}_j(t) - \mathbf{r}_i(t)$ and the relative velocity $\mathbf{v}_{ij} = \mathbf{v}_j(t) - \mathbf{v}_i(t)$ between the colliding particles $i$ and $j$. Specifically, the time until the next collision $t_\mathrm{col}$ is given by \cite{rapaport2004art}
\begin{equation}
    t_\mathrm{col} = \frac{-b - \sqrt{b^2 - v_{ij}^2(r_{ij}^2 - \sigma_{ij}^2)}}{v_{ij}^2}, \label{eq:tcol}
\end{equation}
where $b = \mathbf{r}_{ij}\cdot \mathbf{v}_{ij}$ and the minimum approach distance $\sigma_{ij} = R_i + R_j$ is the sum of the radii $R_{i}, R_{j}$ of the two spheres. Note that a collision is only expected to occur when $b < 0$ (i.e. the particles are currently moving towards each other), and the discriminant $b^2 - v_{ij}^2(r_{ij}^2 - \sigma_{ij}^2) \ge 0$. 

When a collision occurs, the two particles involved will instantaneously undergo a change in their velocities. Assuming a perfectly elastic collision, this change conserves the total momentum and kinetic energy of the particles, resulting in velocity changes:
\begin{equation}
 \delta \mathbf{v}_i = -\frac{2 m_j}{m_i+m_j} \frac{b \mathbf{\hat{r}}_{ij}}{\sigma_{ij}},  
 \hspace{1cm}
 \delta \mathbf{v}_j =  \frac{2 m_i}{m_i+m_j} \frac{b \mathbf{\hat{r}}_{ij}}{\sigma_{ij}^2},  
\end{equation}
where $m_i$ is the mass of particle $i$, $\mathbf{\hat{r}}_{ij}$ is a unit vector in the direction of $\mathbf{r}_{ij}$, and the values of $b$ and $\mathbf{\hat{r}}_{ij}$ are taken at the moment of collision.

Using these equations for predicting and resolving collisions, a naive implementation of the algorithm could simply proceed as follows:
\begin{enumerate}
    \item Check \emph{all} particle pairs for possible collisions, and record the first collision time $t_\mathrm{first}$.
    \item Set $t = t_\mathrm{first}$ and update \emph{all} particle positions.
    \item Process the collision by updating the velocities of the two colliding particles
    \item Go to step 1.
\end{enumerate}
In practice, this is rather inefficient, and there are a number of standard optimizations that can be implemented to improve performance. First, after a collision between particles $i$ and $j$, any other collisions previously predicted for particle $i$ and $j$ are no longer valid, and hence the future collisions for these particles need to be recalculated. However, all predicted collisions that only involve particles \emph{other} than $i$ and $j$ are still valid, since the trajectories of those particles did not change. To make use of this, we keep track of an \emph{event calendar}: a list of predicted future events. Whenever a collision occurs, we then remove any future events that involve the colliding particles from the event calendar, but leave the others in place, avoiding the need to re-predict them. We will discuss the implementation of the event calendar in Sec. \ref{sec:calendar}.

A second useful optimization lies in the fact the positions of any other particles are irrelevant for handling a collision between two particles. Hence, there is no need to update particle positions for any collision they are not involved in. Instead, for each particle $i$ we can simply keep track of the last time $t_i$ that the particle was updated, and its position and velocity at that time, and update its position only when needed. This avoids the vast majority of particle updates in the algorithm described above.

Finally, we can save additional time by realizing that particles can only collide with particles in their immediate environment. Hence, predicting collisions between particles that are separated by distances many times larger than the particle size is rather wasteful: most likely these events will be invalidated by intervening collisions long before they would take place. Efficiently dealing with particles with a finite interaction range is a common problem in computer simulations, and the standard approach for addressing this is to only consider a limited selection of nearby neighbors when calculating interactions or predicting collisions \cite{bookfrenkel}. In EDMD simulations, the most commonly used method is the use of cell lists \cite{rapaport2004art, bannerman2011dynamo}, but neighbor list (also called Verlet lists) have also been implemented \cite{donev2005neighbor}. In both cases, we introduce a new event type, a neighborhood event, which triggers an update of the neighbor list or cell list position for a given particle. These methods will be discussed in more detail in Section \ref{sec:neighbor}.

As a result, the basic outline of a modern EDMD simulation is as follows.
\begin{enumerate}
    \item Initialize the cell list or neighbor lists.
    \item For each particle, predict the point in time when  either it moves to a different cell or its neighbor list needs an update, and add this event to the event calendar.
    \item For all neighboring particle pairs, predict future collisions and add them to the event calendar.
    \item Locate the first collision or neighborhood event in the event calendar, and move the simulation time $t$ forward to the associated event time.
    \item If the current event is a neighborhood event, do the necessary bookkeeping on the particle environment, and predict new neighborhood and collision events for the updated particle.
    \item If the current event is a collision, update the two colliding particles to time $t$, and process the collisions. Then, predict new neighborhood and collision events for both particles.
    \item Go to step 4.
\end{enumerate}
In the following subsections, we look at some of the implementation details in more detail.

\subsection{Event calendar}
\label{sec:calendar}
The implementation of the event calendar in EDMD simulations has been discussed extensively in literature \cite{lubachevsky1991simulate, marin1993efficient, marin1995empirical, rapaport2004art, paul2007complexity}. We are storing a large number of events (scaling with the system size $N$), each containing (at least) information about the event type, the time at which the event will occur, and the particles involved. The data structure we choose for this has to allow for 
\begin{enumerate}
    \item rapidly finding the event that is first chronologically (i.e. with the lowest event time),
    \item efficient addition of new events, and
    \item efficient deletion of invalidated events.
\end{enumerate}
The first requirement strongly suggests that the events in the event calendar should be kept sorted based on their event time. Using an unsorted event calendar, finding the first event would require a search through all stored events ($\bigO{N}$ operations), which would be expensive to perform for every collision. A sorted data structure avoids this issue by ensuring that the next event in line is always easy to find. In particular, event calendars in the form of a \emph{binary search tree} are commonly used \cite{rapaport2004art}, but other variations of binary tree structures have been investigated as well \cite{marin1995empirical}. Regardless of the exact implementation, in binary tree structures the total cost of operations for handling a collision (i.e. the combination of finding the first event, deleting the old events for the particles involved, and inserting the newly predicted collisions) scales as $\bigO{\log N}$ in system size. Since the number of collisions per simulation time  unit in an EDMD trivially scales as $\bigO{N}$, this results in a total computational cost per time unit of the simulation of $\bigO{N \log N}$. 

A more efficient event calendar strategy was introduced in Ref. \onlinecite{paul2007complexity}, which proposed combining a binary tree structure for events in the near future with an array of linear lists for the remaining events. In particular, this strategy divides the simulation time into intervals of length $\Delta t$, and chooses to \emph{only} store events within the current time interval in a sorted binary tree structure as described above. The remaining events are stored inside a set of ``event buckets'', with each bucket corresponding to an unsorted list of all events that occur within the same time interval $t \in [k \Delta t, (k+1)\Delta t)$ (with $k$ an integer). At the end of the current time interval (i.e. when no more events remain in the binary tree), the events from the next event bucket are added to the binary tree, and the simulation continues. Since the events in each bucket are implemented as unsorted doubly linked lists, addition or removal of an event from the buckets can be done at a fixed cost of $\bigO{1}$. Moreover,  the cost of updating the initial binary tree scales only as $\bigO{\log n}$, with $n$ the typical number of events scheduled in a time interval $\Delta t$. By choosing $\Delta t \propto 1/N$, this number $n$ can be chosen to be independent of $N$, resulting in a $\bigO{1}$ cost for event scheduling and an overall EDMD simulation performance of $\bigO{N}$ operations per simulation time unit. This yields a significant performance improvement over implementations that rely on a pure binary tree, especially for larger systems (i.e. $N\gtrsim 10^3$) \cite{paul2007complexity}.

Note that in this approach, care needs to be taken that a sufficiently large number of event buckets are available. The number of buckets should typically be on the order of $t_\mathrm{max} / \Delta t$, where $t_\mathrm{max}$ is the maximum expected difference between the current time and a predicted event time. In practice, in an EDMD simulation this maximum time is not bounded, as two particles that happen to move slowly or nearly parallel can experience a collision arbitrarily far in the future. However, as described in Ref. \onlinecite{paul2007complexity}, this issue can be circumvented by the use of an ``overflow'' bucket which contains the (rare) events that are too far into the future. In all our simulation codes presented here, we follow the algorithm described in Ref. \onlinecite{paul2007complexity} for the event calendar, with one minor modification: we use a binary search tree rather than a complete binary tree \cite{marin1995empirical} as our initial event tree. Since  the scheduling of events into the (small) binary tree takes up only a small fraction of the overall simulation time, we do not expect this choice to make a significant difference.

\subsection{Neighbor selection}
\label{sec:neighbor}

The majority of the computational cost of an EDMD simulation is spent on the prediction and scheduling of collisions. Hence, reducing the number of collision partners to check for a given particle can drastically affect simulation times. Broadly, there are two main strategies for determining which particles to check. 

Most commonly used are \emph{cell lists}: the simulation box is divided up into a grid of cells that are sufficiently large to ensure that collisions can only occur between particles in adjacent cells (i.e. particles that are at most one cell away in along any of the three spatial dimensions). We can then limit ourselves to only predicting collisions with particles that are either in the same cell as the particle being checked, or in any of the 26 (in 3D) adjacent cells. Since the number of particles per cell is limited, finding the future collisions for a given particles then only requires checking a limited number of neighbors, rather than all $N$ particles in the system. This efficiency comes at the (small) cost of having to keep track of the cell each particle is in. To this end, an extra event type is added that occurs when a particle crosses from one cell to the next. Like collisions, these cell-crossing events are predicted and added to the event calendar each time a particle is involved in an event. When a cell crossing occurs, particles in the newly adjacent cells (i.e. 9 cells in 3 dimensions) are checked for collisions.

Another common approach for neighbor selection is via the implementation of \emph{neighbor lists} \cite{bookfrenkel}. An efficient implementation of this approach for event-driven simulations was shown in Ref. \onlinecite{donev2005neighbor}. In this strategy, each particle maintains a list of particles that are considered to be ``close enough'' for possible interactions. In particular, we store for each particle $i$ the center of its current neighborhood $\mathbf{r}_{\mathrm{neigh},i}$, and the set of other particles $j$ such that 
\begin{equation}
\left|\mathbf{r}_{\mathrm{neigh},i} - \mathbf{r}_{\mathrm{neigh},j} \right| < (1+\alpha) \sigma_{ij},
\end{equation}
with $\alpha > 1$ a parameter that controls the size of the neighborhood of a particle. We can then conclude that particle $i$ cannot collide with any particle $k$ \emph{not} in its neighborhood, unless either $i$ has moved by at least $\alpha R_i$ or $k$ has moved by at least $\alpha R_k$ from their respective neighborhood centers. Hence, we can safely limit ourselves to predicting collisions only between particles that are considered to be neighbors, as long as for each particle, we update its neighbor list when it reaches a distance of $\alpha$ times its radius from the center of its environment.

In addition to particle collisions, we then also include an event for each particle that takes place when
\begin{equation}
\left|\mathbf{r}_i - \mathbf{r}_{\mathrm{neigh},i}\right| = \alpha R_i.
\end{equation}
The prediction of this event is analogous to a normal sphere-sphere collision, with an event time
\begin{equation}
  t_\mathrm{neigh} = \frac{-b_{n,i} + \sqrt{b_{n,i}^2 - v_{i}^2(\left|\mathbf{r}_{i} -\mathbf{r}_{\mathrm{neigh},i}\right|^2 - \alpha^2 R_i^2)}}{v_{i}^2},
\end{equation}
with $b_{n,i} = \mathbf{r}_i \cdot \mathbf{r}_{\mathrm{neigh},i}$. The neighbor list update affects not only the list of neighbors of particle $i$, but will also cause $i$ to be added to (or removed from) the neighbor lists of nearby particles. Note that a cell list is still used to find possible neighbors of a given particle, in order to limit the number of distances that need to be checked. The cell size in this case is chosen to be large enough to ensure that only particles in neighboring cells can be in each others neighbor list. Updates to the cell list are done as part of the neighbor list update. 

Analogous to Ref. \cite{donev2005neighbor}, we implement the neighbor list update of particle $i$ as follows:
\begin{enumerate}
    \item Update particle $i$ to the current simulation time.
    \item If particle $i$ has left the simulation box, apply periodic boundaries.
    \item Set $\mathbf{r}_{\mathrm{neigh},i} = \mathbf{r}_{i}$
    \item Update the cell list with the new particle location.
    \item For all particles $j$ that were previously neighbors of $i$, remove $i$ from the neighbor list of $j$ and \emph{vice versa}.
    \item Check all 27 neighboring cells for particles $j$ that satisfy:  $\left|\mathbf{r}_{\mathrm{neigh},j} - \mathbf{r}_{\mathrm{neigh},i}\right| < (1+\alpha)\sigma_{ij}$. For each of these, add $i$ to the neighbor list of $j$ and \emph{vice versa}.
    \item Delete any future events for particle $i$ from the event calendar.
    \item Predict and schedule new collision and neighbor list events for particle $i$.
\end{enumerate}

We implement the neighbor list as a fixed-length array for each particle, allowing a maximum number $n_\mathrm{neigh}^\mathrm{max}$ of neighbors that is constant during the simulation and should be chosen large enough to accommodate the maximum number of neighbors that a particle can have at the chosen shell size $\alpha$. For the hard-sphere systems studied here, we found that a shell size $\alpha = 0.5$ gave near optimal results (see SI). For this choice, $n_\mathrm{neigh}^\mathrm{max} = 24$ was sufficient in all cases \footnote{Note that in principle, it is possible to implement a variation on this approach where the shell size of a specific particle is temporarily reduced if too many neighbors are encountered \cite{donev2005neighbor}, but this was found to be unnecessary for the systems considered here, where a good upper bound on the maximum number of nearest neighbors can be estimated.}.

It should be emphasized that the cell list and neighbor list approaches will perform optimally under different circumstances. In particular, neighbor lists have excellent performance in systems where particles are likely to remain in the same environment for a long time, as is the case in solid phases and glassy fluids. Under these circumstances, the neighbor lists require few updates, and the number of calculated collisions is typically lower than in a cell list approach. For example, in a monodisperse hard-sphere crystal at a number density $\rho \sigma^3 \approx 1$, the neighbor list for each particle is likely to only contain its 12 nearest neighbors, while checking 27 cells of a size of at least $\sigma^3$ each requires checking on the order of 27 particles. In contrast, in low-density fluids, the high particle mobility leads to much more frequent neighbor list updates. As neighbor lists updates are more expensive than
cell list updates, this favors the cell list method. Since long simulation times are more likely to be a concern in systems with relatively slow dynamics (e.g. glassy systems), we focus here on the neighbor list approach, but include benchmarks for one code (labeled \CELL{}) using cell lists.

\subsection{Lazy event scheduling}

A key point of consideration in the implementation of an EDMD simulation code is the question of how many events to schedule per particle. In the above, we have implied that all predicted collisions (and the neighbor list update) for each particle are added to the event tree, but this is not necessarily the most efficient choice. Lubachevsky\cite{lubachevsky1991simulate} noted that when predicting many events per particle, the later events are likely to be pre-empted by the earlier ones, and hence the computational cost of adding them to the event calendar might well be a waste. As a result, many studies (e.g. Refs. \onlinecite{lubachevsky1991simulate, donev2005neighbor, khan2011parallel}) choose to schedule only a single event per particle. If the event scheduled for particle $i$ is invalidated by a collision between two other particles, the event is typically converted into an update event, which simply rechecks for future collisions and neighborhood events without changing the particle trajectory. Although this means that sometimes the same collision is predicted more than once, the reduction in computation cost due to scheduling fewer events can result in significant speedups \cite{khan2011parallel}. An alternative strategy, used in e.g. Refs. \cite{marin1993efficient, bannerman2011dynamo} does maintain all predicted events per particle, but still schedules only one event per particle in the event tree (limiting its size), while separately maintaining lists of sorted events per particle.

Here, we explore both the option of scheduling multiple events per particle and scheduling only a single event per particle. In particular, we consider two EDMD codes, both using neighbor lists and the event scheduling approach from Ref. \onlinecite{paul2007complexity}, with the first (labeled \MULTI{}) scheduling multiple events per particle, and the second (labeled \SINGLE{}) scheduling only one event per particle.

Specifically, in the \MULTI{} code, we schedule all collision events that take place before the next neighbor list update event for both colliding particles. Since our implementation of the neighbor list update of particle $i$ completely reevaluates the future events of $i$, there is no benefit to scheduling any events beyond this time.

In contrast, in the \SINGLE{} code, we only schedule the first (collision or neighbor list update) event for each particle. In order to check for invalidated events, each particle keeps track of the total number of collisions $n_\mathrm{col}$ it has experienced. In particular, when updating the event for particle $i$ (i.e. after a collision or neighbor list update), we do the following:
\begin{enumerate}
    \item Predict the new neighbor list update for particle $i$, and set $t_\mathrm{min}$ equal to the predicted time.
    \item For all neighbors of $i$, predict a possible future collision. If a predicted collision time is smaller than $t_\mathrm{min}$, set $t_\mathrm{min}$ equal to the collision time, and remember the collision partner.
    \item Schedule the new (neighbor list or collision) event for particle $i$ at $t_\mathrm{min}$.
    \item If the predicted event was a collision (with particle $j$), store $n_{\mathrm{col},j}$ as additional information with the event. 
\end{enumerate}
Then, when handling the collision that was predicted for particle $i$ with collision partner $j$, we:
\begin{enumerate}
    \item Remove the event from the event calendar.
    \item Check that the value of $n_{\mathrm{col},j}$ stored with the event predicted for particle $i$ is equal to the current value of $n_{\mathrm{col},j}$.
    \item If not, then the collision is no longer valid. Predict the next event for particle $i$, and move on to the next event (ignore the steps below).
    \item Otherwise, resolve the collision between the two particles, and increment $n_{\mathrm{col},i}$ and $n_{\mathrm{col},j}$.
    \item Remove the scheduled event for $j$ from the event calendar, and predict new events for both particles.
\end{enumerate}

Our approach is similar to the one followed in Ref. \onlinecite{donev2005neighbor}, with the distinction that we make no attempt to ensure that collision predictions are symmetric between collision partners. For example, if the first collision found for particle $i$ is with particle $j$, then this does not affect the predicted collision for particle $j$, regardless of the predicted collision time. Since the $i-j$ collision will be handled when the event associated with particle $i$ is handled, there is no need to also schedule the same event in reverse. Moreover, the original prediction for $j$ may still take place if the $i-j$ collision is invalidated, so it is beneficial to leave the original event prediction for $j$ in place.

In addition to saving significant amounts of scheduling effort and memory usage, this scheduling approach allows for a few small optimizations in the collision prediction. First, the strict time limit on predicted events can sometimes avoid the full calculation of the collision time in Eq. \ref{eq:tcol}, by noting that $t_\mathrm{col}$ is always greater than $ (r_{ij}^2 - \sigma_{ij}^2) / 2b$. Hence, if this lower bound is already greater than the current $t_\mathrm{min}$, there is no need to calculate the exact collision time. Additionally, the total number of events that are scheduled is now equal to the number of particles, plus a small number of special events that are used for e.g. performing measurements, writing simulation output, or applying a thermostat. This eliminates the need for separate data structures for events and particles, allowing us to combine both into a single structure. This may help in optimizing memory access. Finally, the typically small times for predicted events means that the number of event buckets used in the event calendar can be quite limited, reducing memory usage.

\section{Benchmarks}
\label{sec:benchmarks}

We test three versions of our simulation code (\CELL{}, \MULTI{}, and \SINGLE{}) on both monodisperse and bidisperse hard-sphere fluids, and compare their performance to other available EDMD codes \cite{rapaport2004art, bannerman2011dynamo}. Some additional practical details of our simulation codes are described in the Supplemental Information (SI). Initial disordered configurations were created using a variation of the \SINGLE{} code where all particles grow in size during the simulation. This results in a rapid compression from a dilute random configuration to the desired packing fraction. After reaching the desired packing fraction, the system was then equilibrated using the \SINGLE{} code. In order to cut down on the computational cost of equilibration (which can be slow in the dense binary system), all initial configurations for systems larger than $N = 10^5$ particles were created by duplicating a smaller configuration either $2\times2\times2$ times or $3\times3\times3$ times.

For all benchmarks, the simulation is run for at least $10^7$ collisions, and we measure the average number of collisions processed per second, starting the measurement after the initialization of the system. All three simulation codes are implemented in C. Compilation and optimization details are given in the SI. All benchmarks were performed on a workstation containing two 3.3GHz Intel Xeon Gold 6234 CPUs, with 8 CPU cores each, using Ubuntu 20.04 as the operating system . For each benchmark, both CPUs were filled by running 16 identical jobs in parallel. Hyperthreading was disabled. 

\subsection{Monodisperse hard spheres}

\begin{figure}
    \centering
    \includegraphics[width=0.9\linewidth]{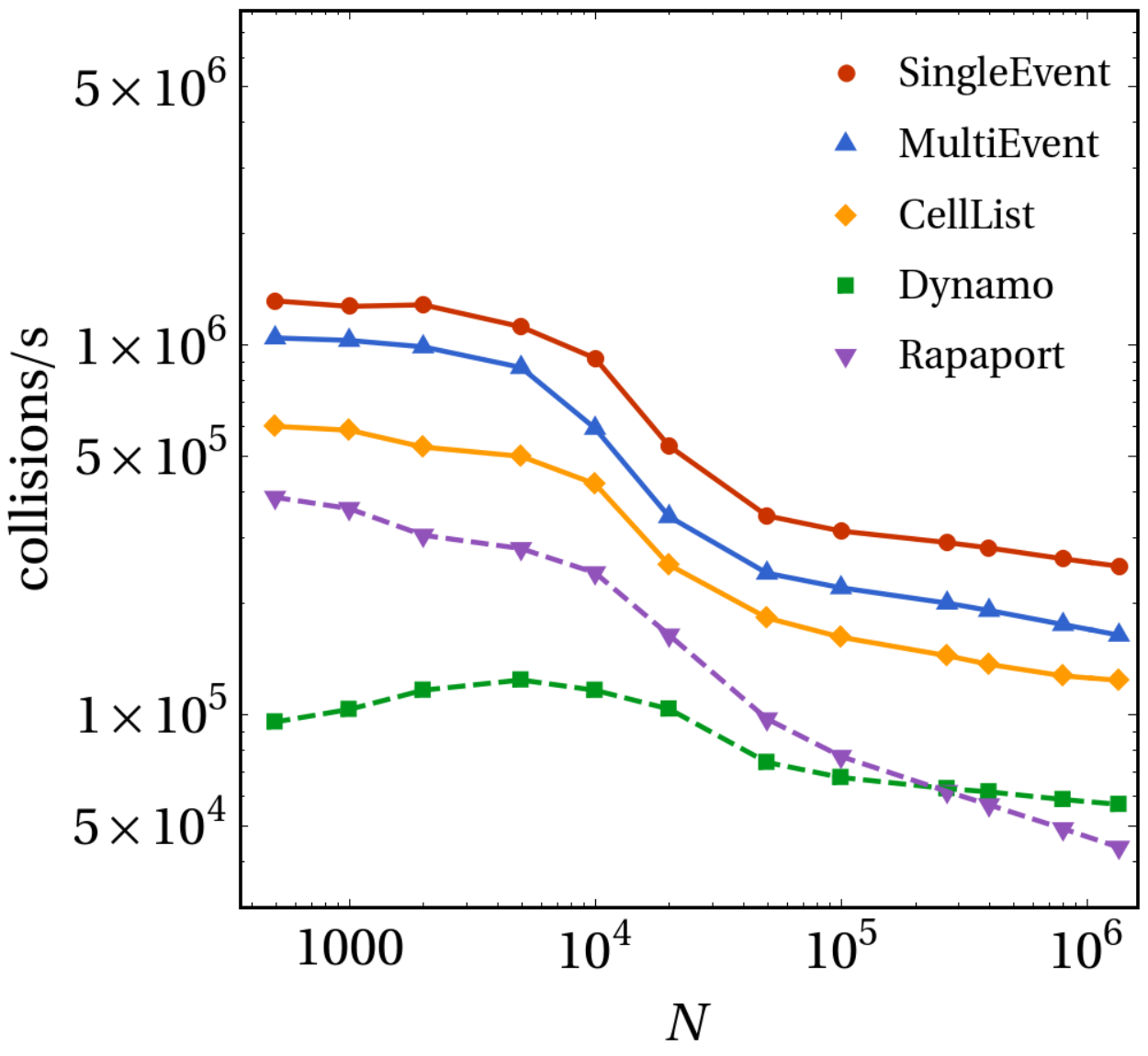}
    \caption{Benchmark results on monodisperse hard-sphere fluids for different simulation codes and a range of system sizes $N$. The label Rapaport indicates the code provided with Ref. \cite{rapaport2004art}. The label Dynamo indicates the simulation package Dynamo \cite{bannerman2011dynamo}.}
    \label{fig:mono}
\end{figure}

First, we simulate a monodisperse hard-sphere fluid at packing fraction $\eta = 0.49$, for a range of system sizes $N$ ranging from 500 to $1.35 \times 10^6$ particles. This packing fraction is slightly below the freezing packing fraction \cite{bookfrenkel}. We compare all three codes to both the code by Rapaport, provided with Ref. \cite{rapaport2004art}\footnote{
Note that small modifications to the initialization and output of the code were made to allow the code to read in an initial configuration and report the elapsed time during the simulation.}, and the simulation package Dynamo developed by Bannerman \emph{et al.}\cite{bannerman2011dynamo}, version 1.7.6. 

In Fig. \ref{fig:mono}, we show the number of collisions per second processed by each simulation. We observe a significant variation in simulation speeds, which is largely consistent as a function of system size. It is interesting to note that even for this relatively mobile hard-sphere fluid, the introduction of a neighbor list (i.e. switching from the \CELL{} to \MULTI{} code) results in a speedup of around $50\%$ accross all system sizes. Combining this with the choice to store only one event per particle (i.e. the \SINGLE{} code) provides another similar speedup. In total, the \SINGLE{} code outperforms Dynamo by a factor 4-10, depending on system size. 

All benchmarked simulations show a noticeable decrease in efficiency around a system size of $\approx 10^4$, which we attribute to cache misses. The simulation methods examined here typically use a few hundred bytes of memory per particle, and the total cache size of the used processors is approximately 2.5MB per core. This means that for small systems (up to a few thousand particles), all data relevant to the simulation fits inside the processor's cache. For larger systems significant data exchange between CPU and the main memory is needed, which drastically slows down performance. As some of the CPU cache is shared between cores, this performance drop can be shifted to slightly larger system sizes by running only a single job at a time (see SI). Note that in EDMD simulations, particle data is typically accessed in an essentially random order: two consecutive collisions can take place anywhere in the system, and particles that are close together in space may not be stored close together in memory. As a result, optimizing EDMD simulations with respect to cache usage is inherently difficult. 

As Fig. \ref{fig:mono} shows, the Rapaport code shows a stronger decrease in performance for large systems than the other methods. This is due to its use of a $\bigO{\log N}$ binary search tree as the event calendar, in contrast to the $\bigO{1}$ scheduling methods used by the other codes.

\subsection{Bidisperse hard spheres}

As our second test system, we take a binary hard-sphere system with a size ratio $q=0.85$ and a composition $x_L = N_L/N = 0.3$ (where $N_L$ is the number of large spheres in the system). This system has been explored in several studies of glassy dynamics \cite{marin2020tetrahedrality, boattini2020autonomously, boattini2021averaging}. Note that since the Rapaport code\cite{rapaport2004art} implements only monodisperse systems, we did not include it in the comparison here. 

\begin{figure}
    \centering
    \includegraphics[width=0.9\linewidth]{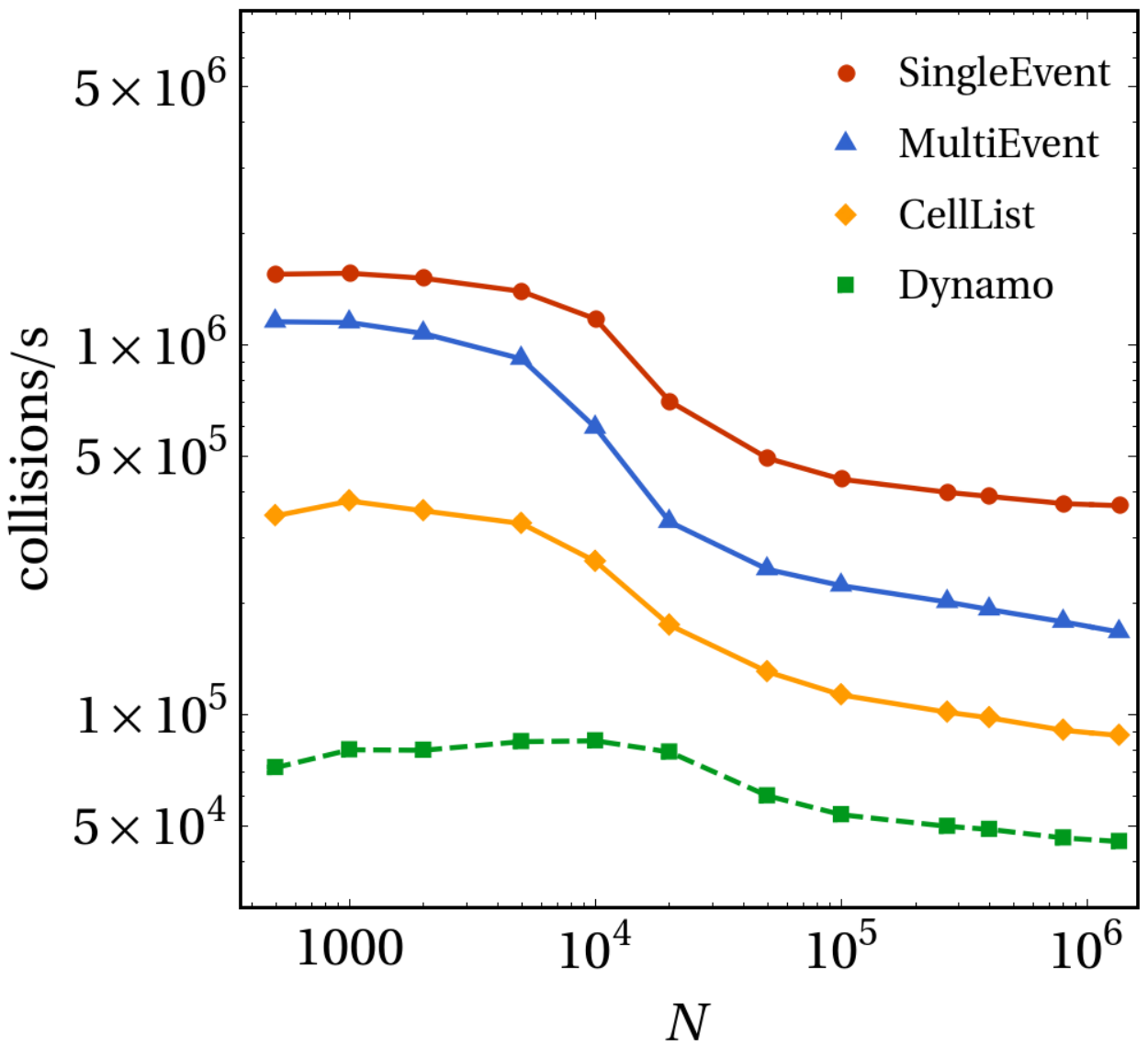}
    \caption{Benchmark results on bidisperse hard-sphere fluids with size ratio $q=0.85$, composition $x_L = 0.3$, and packing fraction $\eta = 0.58$, as a function of the system size $N$.}
    \label{fig:binary_sizes}
\end{figure}

For this binary mixture, we first benchmark the simulation performance as a function of system size at a fixed packing fraction $\eta = 0.58$. The results are shown in Fig. \ref{fig:binary_sizes}. We observe similar results as for the monodisperse system, with again significant performance gaps between the different simulation codes. In this system, dynamics are glassy: particles are trapped by their neighbors for significant amounts of time before escaping their cage. As a result, neighbor updates are relatively rare, and hence the effect of incorporating a neighbor list is more significant than in Fig. \ref{fig:mono}: it results in a speed-up of approximately a factor two. Additionally, the typical number of events predicted per particle in the \MULTI{} code is higher in this system than in the lower-density monodisperse system, resulting in an increased performance boost from switching to the \SINGLE{} code.

\begin{figure}
    \centering
    \includegraphics[width=0.9\linewidth]{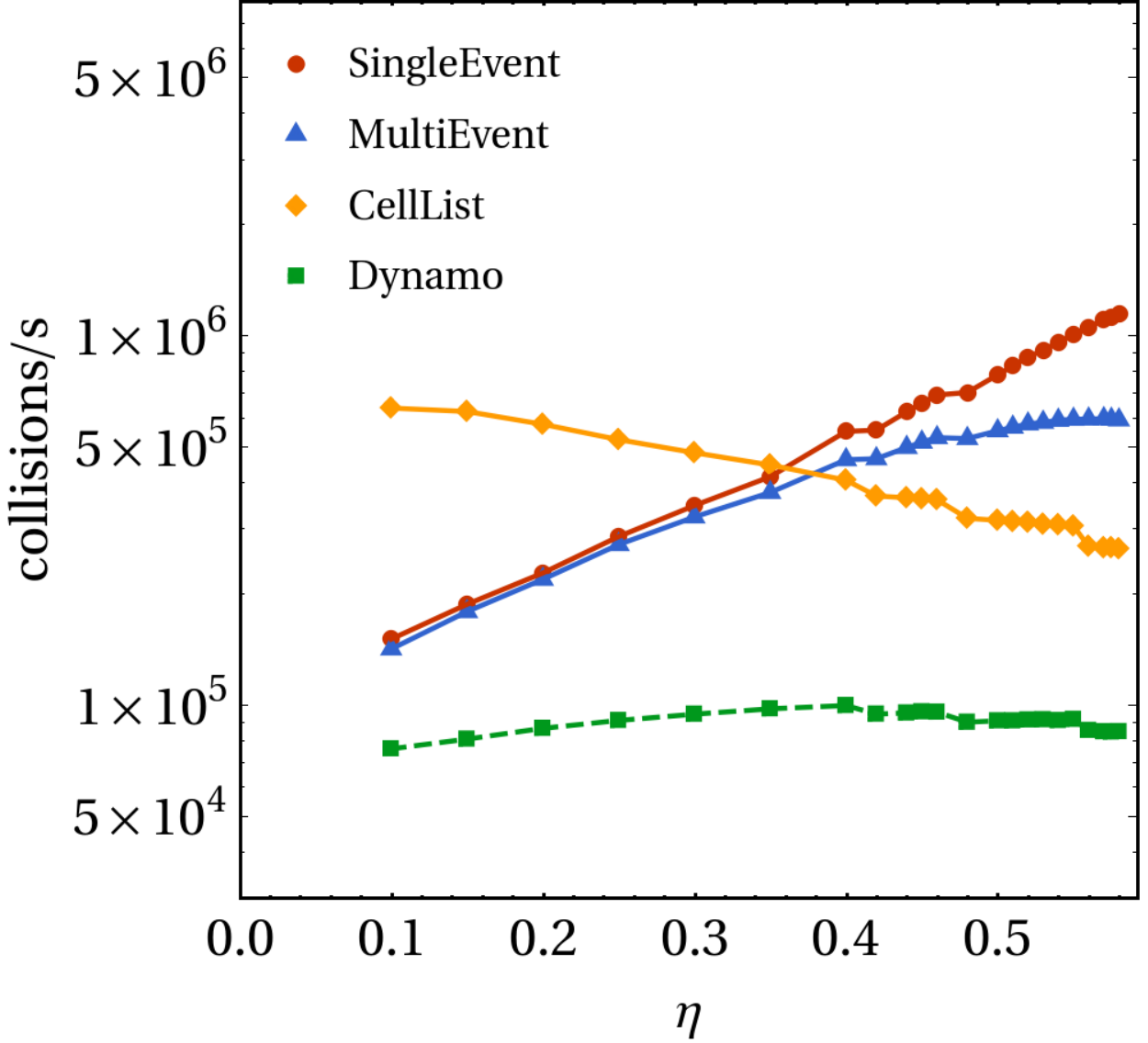}
    \caption{Benchmark results on bidisperse hard-sphere fluids with size ratio $q=0.85$, composition $x_L = 0.3$, and system size $N=10^4$, as a function of the packing fraction $\eta$.}
    \label{fig:binary_pf}
\end{figure}

In order to investigate the effect of packing fraction on performance, we also benchmark each simulation code at a fixed system size of $N=10^4$ particles for a range of different packing fractions. As shown in Fig. \ref{fig:binary_pf}, we find a significant variation in the efficiency of the different codes as the packing fraction changes. As already hinted at in Sec. \ref{sec:neighbor}, the \CELL{} code performs excellently at low packing fractions, but slows down as the packing fraction increases. This is understandable: as the packing fraction increases, more and more particles have to be checked for collisions. As most of these events will be pre-empted by an earlier collision, many of these calculations are wasted, resulting in an overall slowdown with increasing packing fraction. The step-wise performance decreases that can be seen for both the \CELL{} code and Dynamo in Fig. \ref{fig:binary_pf} are due to the discrete reduction the number of cells that fit in the simulation box. A step occurs at a packing fraction where the number of cells is reduced by 1, which results in a discontinuous increase in the average number of particles per cell, and hence an increase in the computation cost of checking for collisions. Similar steps can be seen even for the \MULTI{} and \SINGLE{} codes, but only at lower packing fractions ($\eta \lesssim 0.5$) where neighbor list updates (which rely on the cell list) account for a significant part of the simulation time.

In contrast to the \CELL{} code, in terms of collisions per second, the neighbor-list based codes become \emph{faster} with increasing packing fraction. As the packing fraction increases, particle mobility decreases and hence fewer neighbor list updates are required\footnote{
    Note that this effect does not hold for the cell list updates in the \CELL{} code: for a homogeneous system, the average (bidirectional) flux of particles through any cell wall is only a function of the particle density and the typical speed (i.e. temperature) of the particles. While some particles may be trapped for a long time in a single cell, this is compensated by other particles trapped vibrating around the edge of a cell and hence undergoing many cell crossings.}.
As a result, both neighbor-list driven codes outperform the \CELL{} code at packing fractions $\eta \gtrsim 0.4$. Moreover, at higher packing fractions the number of unnecessary events predicted scheduled per particle in the \MULTI{} code is larger, and hence the \SINGLE{} code becomes the fastest option by a significant margin. Note that most hard-sphere simulation studies where long simulations are required are likely to focus on the high-packing fraction fluid regime, where e.g. glassy behavior or spontaneous crystal nucleation can be observed.

\section{Discussion and conclusions}
\label{sec:discuss}

The simulation codes presented here (and in particular the \SINGLE{} code) clearly perform well in comparison to other publicly available EDMD implementations. It should be noted that part of this efficiency is due to the fact that the code was  written with only hard-sphere systems in mind. In contrast, Dynamo\cite{bannerman2011dynamo} offers the possibility to simulate different particle shapes, different boundary conditions, stepped pair interactions, as well as a variety of other features. However, for the common use case of studying hard-sphere systems in a periodic rectangular box, as is often necessary for e.g. glass or nucleation studies, the presented  difference in simulation speed (typically a factor 5 to 10) may be a crucial factor, saving significantly on both CPU time and energy consumption. 

It should be noted that if the goal is simply to equilibrate a hard-sphere system or to sample equilibrium states, it is not \emph{a priori} clear that EDMD simulations are the ideal choice. Recent work by Klement \emph{et al.} demonstrated that hard-sphere systems can be highly efficiently equilibrated using Newtonian Event Chains (NEC) \cite{klement2019efficient}, an approach that blends aspects of Monte Carlo simulations and event-driven dynamics. Simulating a monodisperse system of hard spheres at $\eta = 0.49$, they observed that their NEC simulations were significantly faster than EDMD simulations in terms of particle displacements per hour, which directly translated into a similar speedup in terms of the simulation time required to simulate e.g. particle diffusion or crystal melting. In Table \ref{tab:NEC}, we compare the number of displacements per second achieved by the currently available implementation of NEC \cite{NECgithub} in HOOMD\cite{anderson2020hoomd} to the number of collisions per second of the \SINGLE{} code for several system sizes. We see that a sufficiently efficient EDMD code can indeed slightly outperform this implementation of NEC
    \footnote{It should be noted that the number of NEC displacements per second in Table \ref{tab:NEC} is significantly lower than those reported in Ref. \onlinecite{klement2019efficient}, especially for larger systems. We attribute this to our protocol of filling up the CPU with an identical number of jobs when performing benchmarks. When running only a single job per CPU, we observe performance in line with Ref. \onlinecite{klement2019efficient}, with EDMD still outperforming NEC in terms of displacements per second.}. 
However, it should be noted that since the complexity of an NEC displacement is lower than that of an EDMD collision, further optimization of the NEC implementation may be possible. Moreover, as also noted in Ref. \onlinecite{klement2019efficient}, the local nature of the NEC algorithm allows for more straightforward parallelization \cite{klement2021newtonian}. On the other hand, the EDMD approach has the significant advantage of producing physically realistic dynamics.

Although here we focus purely on single-processor simulation codes, simulations of large systems can often be parallelized to decrease simulation time, although this typically carries an overhead cost in total CPU time. For Monte Carlo methods, highly parallelized approaches have been shown to be capable of extremely  fast equilibration of 2D hard-disk systems \cite{anderson2013massively, engel2013hard}. A similar approach may be effective for 3D hard spheres as well. In contrast, EDMD simulations are notoriously difficult to parallelize \cite{miller2004event}, due to the fact that two events that are spatially far apart could influence each other  even if they occur at nearly the same time, as long as they are connected by a chain of (nearly) touching particles. Nonetheless, several effective methods for parallelization of EDMD simulations have been demonstrated (see e.g. \cite{miller2004event, khan2011parallel}), and can in principle be extended to the EDMD codes presented here. 

\begin{table}
\centering
    \begin{tabular}{r|c|c}
    N     &  NEC displacements/s & EDMD collisions/s\\
    % \hline
    \numprint{1000}     & $1.0 \times 10^6$ & $1.6 \times 10^6$ \\
    \numprint{10000}    & $8.8 \times 10^5$ & $1.2 \times 10^6$ \\
    \numprint{100000}   & $4.1 \times 10^5$ & $4.3 \times 10^5$ \\
    \numprint{1350000}  & $2.2 \times 10^5$ & $3.7 \times 10^5$ \\
    % \hline
    \end{tabular}
    \caption{Comparison between Newtonian Event Chain (NEC) simulations \cite{klement2019efficient, NECgithub}, and the \SINGLE{} EDMD code, for a selection of the systems shown in Fig. \ref{fig:binary_sizes}. Following Ref. \onlinecite{klement2019efficient}, we compare the number NEC particle displacements per second to EDMD collisions per second. }
    \label{tab:NEC}
\end{table}

As a final note, it is interesting that in all of the benchmarks, we found that the \SINGLE{} code performed at least as efficiently as the \MULTI{} code. This observation is likely to be general for hard spheres and any extensions of the code to other spherically symmetric systems. In these systems, the cost of predicting a collision is low, and hence the cost of having to sometimes recalculate the same prediction is acceptable. However, it is worth noting that in EDMD simulations of anisotropic particles, such as polyhedra, ellipsoids, or patchy particles \cite{donev2005neighbor,donev2005neighbor2, hernandez2007discontinuous, smallenburg2012vacancy, smallenburg2013liquids, marin2019slowing}, the computational cost of predicting a collision event is typically much higher, since collisions for these particles cannot be predicted analytically. This may favor an approach where multiple events are stored per particle, to avoid repeated predictions of the same event. On the other hand, the restriction to a single event per particle also helps to limit the time window to search for possible collisions, hence providing an efficient ``early out'' for numerical collision-prediction routines. Hence, the benefit of single-event scheduling approaches may depend on the system under consideration.

In summary, we have presented a set of light-weight EDMD simulation codes for hard-sphere systems that appear to perform very efficiently in comparison to the the simulations currently used in literature on e.g. the glassy dynamics of hard spheres. Simulating these systems more efficiently allows for a significant saving of CPU time and energy. 

The simulation codes are freely available at:\\ 
\href{https://github.com/FSmallenburg/EDMD}{https://github.com/FSmallenburg/EDMD}

\section{Acknowledgements}
I would like to thank Michael Engel, Marco Klement, and Joshua Anderson for helpful discussions and aid in the comparison to the Newtonian Event Chain algorithm, and Laura Filion for many useful discussions.

%\bibliography{refs}

\begin{thebibliography}{69}%
\makeatletter
\providecommand \@ifxundefined [1]{%
 \@ifx{#1\undefined}
}%
\providecommand \@ifnum [1]{%
 \ifnum #1\expandafter \@firstoftwo
 \else \expandafter \@secondoftwo
 \fi
}%
\providecommand \@ifx [1]{%
 \ifx #1\expandafter \@firstoftwo
 \else \expandafter \@secondoftwo
 \fi
}%
\providecommand \natexlab [1]{#1}%
\providecommand \enquote  [1]{``#1''}%
\providecommand \bibnamefont  [1]{#1}%
\providecommand \bibfnamefont [1]{#1}%
\providecommand \citenamefont [1]{#1}%
\providecommand \href@noop [0]{\@secondoftwo}%
\providecommand \href [0]{\begingroup \@sanitize@url \@href}%
\providecommand \@href[1]{\@@startlink{#1}\@@href}%
\providecommand \@@href[1]{\endgroup#1\@@endlink}%
\providecommand \@sanitize@url [0]{\catcode `\\12\catcode `\$12\catcode
  `\&12\catcode `\#12\catcode `\^12\catcode `\_12\catcode `\%12\relax}%
\providecommand \@@startlink[1]{}%
\providecommand \@@endlink[0]{}%
\providecommand \url  [0]{\begingroup\@sanitize@url \@url }%
\providecommand \@url [1]{\endgroup\@href {#1}{\urlprefix }}%
\providecommand \urlprefix  [0]{URL }%
\providecommand \Eprint [0]{\href }%
\providecommand \doibase [0]{http://dx.doi.org/}%
\providecommand \selectlanguage [0]{\@gobble}%
\providecommand \bibinfo  [0]{\@secondoftwo}%
\providecommand \bibfield  [0]{\@secondoftwo}%
\providecommand \translation [1]{[#1]}%
\providecommand \BibitemOpen [0]{}%
\providecommand \bibitemStop [0]{}%
\providecommand \bibitemNoStop [0]{.\EOS\space}%
\providecommand \EOS [0]{\spacefactor3000\relax}%
\providecommand \BibitemShut  [1]{\csname bibitem#1\endcsname}%
\let\auto@bib@innerbib\@empty
%</preamble>
\bibitem [{\citenamefont {Pusey}\ and\ \citenamefont
  {Van~Megen}(1986)}]{pusey1986phase}%
  \BibitemOpen
  \bibfield  {author} {\bibinfo {author} {\bibfnamefont {P.~N.}\ \bibnamefont
  {Pusey}}\ and\ \bibinfo {author} {\bibfnamefont {W.}~\bibnamefont
  {Van~Megen}},\ }\href@noop {} {\bibfield  {journal} {\bibinfo  {journal}
  {Nature}\ }\textbf {\bibinfo {volume} {320}},\ \bibinfo {pages} {340}
  (\bibinfo {year} {1986})}\BibitemShut {NoStop}%
\bibitem [{\citenamefont {Kegel}\ and\ \citenamefont {van
  Blaaderen}(2000)}]{kegel2000direct}%
  \BibitemOpen
  \bibfield  {author} {\bibinfo {author} {\bibfnamefont {W.~K.}\ \bibnamefont
  {Kegel}}\ and\ \bibinfo {author} {\bibfnamefont {A.}~\bibnamefont {van
  Blaaderen}},\ }\href@noop {} {\bibfield  {journal} {\bibinfo  {journal}
  {Science}\ }\textbf {\bibinfo {volume} {287}},\ \bibinfo {pages} {290}
  (\bibinfo {year} {2000})}\BibitemShut {NoStop}%
\bibitem [{\citenamefont {Royall}\ \emph {et~al.}(2013)\citenamefont {Royall},
  \citenamefont {Poon},\ and\ \citenamefont {Weeks}}]{royall2013search}%
  \BibitemOpen
  \bibfield  {author} {\bibinfo {author} {\bibfnamefont {C.~P.}\ \bibnamefont
  {Royall}}, \bibinfo {author} {\bibfnamefont {W.~C.}\ \bibnamefont {Poon}}, \
  and\ \bibinfo {author} {\bibfnamefont {E.~R.}\ \bibnamefont {Weeks}},\
  }\href@noop {} {\bibfield  {journal} {\bibinfo  {journal} {Soft Matter}\
  }\textbf {\bibinfo {volume} {9}},\ \bibinfo {pages} {17} (\bibinfo {year}
  {2013})}\BibitemShut {NoStop}%
\bibitem [{\citenamefont {Frenkel}\ and\ \citenamefont
  {Ladd}(1984)}]{frenkel1984new}%
  \BibitemOpen
  \bibfield  {author} {\bibinfo {author} {\bibfnamefont {D.}~\bibnamefont
  {Frenkel}}\ and\ \bibinfo {author} {\bibfnamefont {A.~J.}\ \bibnamefont
  {Ladd}},\ }\href@noop {} {\bibfield  {journal} {\bibinfo  {journal} {The
  Journal of chemical physics}\ }\textbf {\bibinfo {volume} {81}},\ \bibinfo
  {pages} {3188} (\bibinfo {year} {1984})}\BibitemShut {NoStop}%
\bibitem [{\citenamefont {Biben}\ and\ \citenamefont
  {Hansen}(1991)}]{biben1991phase}%
  \BibitemOpen
  \bibfield  {author} {\bibinfo {author} {\bibfnamefont {T.}~\bibnamefont
  {Biben}}\ and\ \bibinfo {author} {\bibfnamefont {J.-P.}\ \bibnamefont
  {Hansen}},\ }\href@noop {} {\bibfield  {journal} {\bibinfo  {journal}
  {Physical review letters}\ }\textbf {\bibinfo {volume} {66}},\ \bibinfo
  {pages} {2215} (\bibinfo {year} {1991})}\BibitemShut {NoStop}%
\bibitem [{\citenamefont {Dijkstra}\ \emph {et~al.}(1998)\citenamefont
  {Dijkstra}, \citenamefont {van Roij},\ and\ \citenamefont
  {Evans}}]{dijkstra1998phase}%
  \BibitemOpen
  \bibfield  {author} {\bibinfo {author} {\bibfnamefont {M.}~\bibnamefont
  {Dijkstra}}, \bibinfo {author} {\bibfnamefont {R.}~\bibnamefont {van Roij}},
  \ and\ \bibinfo {author} {\bibfnamefont {R.}~\bibnamefont {Evans}},\
  }\href@noop {} {\bibfield  {journal} {\bibinfo  {journal} {Physical review
  letters}\ }\textbf {\bibinfo {volume} {81}},\ \bibinfo {pages} {2268}
  (\bibinfo {year} {1998})}\BibitemShut {NoStop}%
\bibitem [{\citenamefont {Dijkstra}(1998)}]{dijkstra1998nonadditive}%
  \BibitemOpen
  \bibfield  {author} {\bibinfo {author} {\bibfnamefont {M.}~\bibnamefont
  {Dijkstra}},\ }\href@noop {} {\bibfield  {journal} {\bibinfo  {journal}
  {Physical Review E}\ }\textbf {\bibinfo {volume} {58}},\ \bibinfo {pages}
  {7523} (\bibinfo {year} {1998})}\BibitemShut {NoStop}%
\bibitem [{\citenamefont {Dijkstra}\ \emph {et~al.}(1999)\citenamefont
  {Dijkstra}, \citenamefont {van Roij},\ and\ \citenamefont
  {Evans}}]{dijkstra1999phase}%
  \BibitemOpen
  \bibfield  {author} {\bibinfo {author} {\bibfnamefont {M.}~\bibnamefont
  {Dijkstra}}, \bibinfo {author} {\bibfnamefont {R.}~\bibnamefont {van Roij}},
  \ and\ \bibinfo {author} {\bibfnamefont {R.}~\bibnamefont {Evans}},\
  }\href@noop {} {\bibfield  {journal} {\bibinfo  {journal} {Physical Review
  E}\ }\textbf {\bibinfo {volume} {59}},\ \bibinfo {pages} {5744} (\bibinfo
  {year} {1999})}\BibitemShut {NoStop}%
\bibitem [{\citenamefont {Filion}\ \emph {et~al.}(2011)\citenamefont {Filion},
  \citenamefont {Hermes}, \citenamefont {Ni}, \citenamefont {Vermolen},
  \citenamefont {Kuijk}, \citenamefont {Christova}, \citenamefont
  {Stiefelhagen}, \citenamefont {Vissers}, \citenamefont {Van~Blaaderen},\ and\
  \citenamefont {Dijkstra}}]{filion2011self}%
  \BibitemOpen
  \bibfield  {author} {\bibinfo {author} {\bibfnamefont {L.}~\bibnamefont
  {Filion}}, \bibinfo {author} {\bibfnamefont {M.}~\bibnamefont {Hermes}},
  \bibinfo {author} {\bibfnamefont {R.}~\bibnamefont {Ni}}, \bibinfo {author}
  {\bibfnamefont {E.}~\bibnamefont {Vermolen}}, \bibinfo {author}
  {\bibfnamefont {A.}~\bibnamefont {Kuijk}}, \bibinfo {author} {\bibfnamefont
  {C.}~\bibnamefont {Christova}}, \bibinfo {author} {\bibfnamefont
  {J.}~\bibnamefont {Stiefelhagen}}, \bibinfo {author} {\bibfnamefont
  {T.}~\bibnamefont {Vissers}}, \bibinfo {author} {\bibfnamefont
  {A.}~\bibnamefont {Van~Blaaderen}}, \ and\ \bibinfo {author} {\bibfnamefont
  {M.}~\bibnamefont {Dijkstra}},\ }\href@noop {} {\bibfield  {journal}
  {\bibinfo  {journal} {Physical review letters}\ }\textbf {\bibinfo {volume}
  {107}},\ \bibinfo {pages} {168302} (\bibinfo {year} {2011})}\BibitemShut
  {NoStop}%
\bibitem [{\citenamefont {Bommineni}\ \emph {et~al.}(2019)\citenamefont
  {Bommineni}, \citenamefont {Varela-Rosales}, \citenamefont {Klement},\ and\
  \citenamefont {Engel}}]{bommineni2019complex}%
  \BibitemOpen
  \bibfield  {author} {\bibinfo {author} {\bibfnamefont {P.~K.}\ \bibnamefont
  {Bommineni}}, \bibinfo {author} {\bibfnamefont {N.~R.}\ \bibnamefont
  {Varela-Rosales}}, \bibinfo {author} {\bibfnamefont {M.}~\bibnamefont
  {Klement}}, \ and\ \bibinfo {author} {\bibfnamefont {M.}~\bibnamefont
  {Engel}},\ }\href@noop {} {\bibfield  {journal} {\bibinfo  {journal}
  {Physical review letters}\ }\textbf {\bibinfo {volume} {122}},\ \bibinfo
  {pages} {128005} (\bibinfo {year} {2019})}\BibitemShut {NoStop}%
\bibitem [{\citenamefont {Bommineni}\ \emph {et~al.}(2020)\citenamefont
  {Bommineni}, \citenamefont {Klement},\ and\ \citenamefont
  {Engel}}]{bommineni2020spontaneous}%
  \BibitemOpen
  \bibfield  {author} {\bibinfo {author} {\bibfnamefont {P.~K.}\ \bibnamefont
  {Bommineni}}, \bibinfo {author} {\bibfnamefont {M.}~\bibnamefont {Klement}},
  \ and\ \bibinfo {author} {\bibfnamefont {M.}~\bibnamefont {Engel}},\
  }\href@noop {} {\bibfield  {journal} {\bibinfo  {journal} {Physical Review
  Letters}\ }\textbf {\bibinfo {volume} {124}},\ \bibinfo {pages} {218003}
  (\bibinfo {year} {2020})}\BibitemShut {NoStop}%
\bibitem [{\citenamefont {Schmidt}\ and\ \citenamefont
  {L{\"o}wen}(1997)}]{schmidt1997phase}%
  \BibitemOpen
  \bibfield  {author} {\bibinfo {author} {\bibfnamefont {M.}~\bibnamefont
  {Schmidt}}\ and\ \bibinfo {author} {\bibfnamefont {H.}~\bibnamefont
  {L{\"o}wen}},\ }\href@noop {} {\bibfield  {journal} {\bibinfo  {journal}
  {Physical Review E}\ }\textbf {\bibinfo {volume} {55}},\ \bibinfo {pages}
  {7228} (\bibinfo {year} {1997})}\BibitemShut {NoStop}%
\bibitem [{\citenamefont {Fortini}\ and\ \citenamefont
  {Dijkstra}(2006)}]{fortini2006phase}%
  \BibitemOpen
  \bibfield  {author} {\bibinfo {author} {\bibfnamefont {A.}~\bibnamefont
  {Fortini}}\ and\ \bibinfo {author} {\bibfnamefont {M.}~\bibnamefont
  {Dijkstra}},\ }\href@noop {} {\bibfield  {journal} {\bibinfo  {journal}
  {Journal of Physics: Condensed Matter}\ }\textbf {\bibinfo {volume} {18}},\
  \bibinfo {pages} {L371} (\bibinfo {year} {2006})}\BibitemShut {NoStop}%
\bibitem [{\citenamefont {Gordillo}\ \emph {et~al.}(2006)\citenamefont
  {Gordillo}, \citenamefont {Mart{\'\i}nez-Haya},\ and\ \citenamefont
  {Romero-Enrique}}]{gordillo2006freezing}%
  \BibitemOpen
  \bibfield  {author} {\bibinfo {author} {\bibfnamefont {M.}~\bibnamefont
  {Gordillo}}, \bibinfo {author} {\bibfnamefont {B.}~\bibnamefont
  {Mart{\'\i}nez-Haya}}, \ and\ \bibinfo {author} {\bibfnamefont {J.~M.}\
  \bibnamefont {Romero-Enrique}},\ }\href@noop {} {\bibfield  {journal}
  {\bibinfo  {journal} {The Journal of chemical physics}\ }\textbf {\bibinfo
  {volume} {125}},\ \bibinfo {pages} {144702} (\bibinfo {year}
  {2006})}\BibitemShut {NoStop}%
\bibitem [{\citenamefont {O{\u{g}}uz}\ \emph {et~al.}(2012)\citenamefont
  {O{\u{g}}uz}, \citenamefont {Marechal}, \citenamefont {Ramiro-Manzano},
  \citenamefont {Rodriguez}, \citenamefont {Messina}, \citenamefont
  {Meseguer},\ and\ \citenamefont {L{\"o}wen}}]{ouguz2012packing}%
  \BibitemOpen
  \bibfield  {author} {\bibinfo {author} {\bibfnamefont {E.~C.}\ \bibnamefont
  {O{\u{g}}uz}}, \bibinfo {author} {\bibfnamefont {M.}~\bibnamefont
  {Marechal}}, \bibinfo {author} {\bibfnamefont {F.}~\bibnamefont
  {Ramiro-Manzano}}, \bibinfo {author} {\bibfnamefont {I.}~\bibnamefont
  {Rodriguez}}, \bibinfo {author} {\bibfnamefont {R.}~\bibnamefont {Messina}},
  \bibinfo {author} {\bibfnamefont {F.~J.}\ \bibnamefont {Meseguer}}, \ and\
  \bibinfo {author} {\bibfnamefont {H.}~\bibnamefont {L{\"o}wen}},\ }\href@noop
  {} {\bibfield  {journal} {\bibinfo  {journal} {Physical review letters}\
  }\textbf {\bibinfo {volume} {109}},\ \bibinfo {pages} {218301} (\bibinfo
  {year} {2012})}\BibitemShut {NoStop}%
\bibitem [{\citenamefont {De~Nijs}\ \emph {et~al.}(2015)\citenamefont
  {De~Nijs}, \citenamefont {Dussi}, \citenamefont {Smallenburg}, \citenamefont
  {Meeldijk}, \citenamefont {Groenendijk}, \citenamefont {Filion},
  \citenamefont {Imhof}, \citenamefont {Van~Blaaderen},\ and\ \citenamefont
  {Dijkstra}}]{de2015entropy}%
  \BibitemOpen
  \bibfield  {author} {\bibinfo {author} {\bibfnamefont {B.}~\bibnamefont
  {De~Nijs}}, \bibinfo {author} {\bibfnamefont {S.}~\bibnamefont {Dussi}},
  \bibinfo {author} {\bibfnamefont {F.}~\bibnamefont {Smallenburg}}, \bibinfo
  {author} {\bibfnamefont {J.~D.}\ \bibnamefont {Meeldijk}}, \bibinfo {author}
  {\bibfnamefont {D.~J.}\ \bibnamefont {Groenendijk}}, \bibinfo {author}
  {\bibfnamefont {L.}~\bibnamefont {Filion}}, \bibinfo {author} {\bibfnamefont
  {A.}~\bibnamefont {Imhof}}, \bibinfo {author} {\bibfnamefont
  {A.}~\bibnamefont {Van~Blaaderen}}, \ and\ \bibinfo {author} {\bibfnamefont
  {M.}~\bibnamefont {Dijkstra}},\ }\href@noop {} {\bibfield  {journal}
  {\bibinfo  {journal} {Nature materials}\ }\textbf {\bibinfo {volume} {14}},\
  \bibinfo {pages} {56} (\bibinfo {year} {2015})}\BibitemShut {NoStop}%
\bibitem [{\citenamefont {Jung}\ and\ \citenamefont
  {Petersen}(2020)}]{jung2020confinement}%
  \BibitemOpen
  \bibfield  {author} {\bibinfo {author} {\bibfnamefont {G.}~\bibnamefont
  {Jung}}\ and\ \bibinfo {author} {\bibfnamefont {C.~F.}\ \bibnamefont
  {Petersen}},\ }\href@noop {} {\bibfield  {journal} {\bibinfo  {journal}
  {Physical Review Research}\ }\textbf {\bibinfo {volume} {2}},\ \bibinfo
  {pages} {033207} (\bibinfo {year} {2020})}\BibitemShut {NoStop}%
\bibitem [{\citenamefont {Wang}\ \emph {et~al.}(2021)\citenamefont {Wang},
  \citenamefont {Dasgupta}, \citenamefont {van~der Wee}, \citenamefont
  {Zanaga}, \citenamefont {Altantzis}, \citenamefont {Wu}, \citenamefont
  {Coli}, \citenamefont {Murray}, \citenamefont {Bals}, \citenamefont
  {Dijkstra} \emph {et~al.}}]{wang2021binary}%
  \BibitemOpen
  \bibfield  {author} {\bibinfo {author} {\bibfnamefont {D.}~\bibnamefont
  {Wang}}, \bibinfo {author} {\bibfnamefont {T.}~\bibnamefont {Dasgupta}},
  \bibinfo {author} {\bibfnamefont {E.~B.}\ \bibnamefont {van~der Wee}},
  \bibinfo {author} {\bibfnamefont {D.}~\bibnamefont {Zanaga}}, \bibinfo
  {author} {\bibfnamefont {T.}~\bibnamefont {Altantzis}}, \bibinfo {author}
  {\bibfnamefont {Y.}~\bibnamefont {Wu}}, \bibinfo {author} {\bibfnamefont
  {G.~M.}\ \bibnamefont {Coli}}, \bibinfo {author} {\bibfnamefont {C.~B.}\
  \bibnamefont {Murray}}, \bibinfo {author} {\bibfnamefont {S.}~\bibnamefont
  {Bals}}, \bibinfo {author} {\bibfnamefont {M.}~\bibnamefont {Dijkstra}},
  \emph {et~al.},\ }\href@noop {} {\bibfield  {journal} {\bibinfo  {journal}
  {Nature Physics}\ }\textbf {\bibinfo {volume} {17}},\ \bibinfo {pages} {128}
  (\bibinfo {year} {2021})}\BibitemShut {NoStop}%
\bibitem [{\citenamefont {Auer}\ and\ \citenamefont
  {Frenkel}(2001)}]{auer2001prediction}%
  \BibitemOpen
  \bibfield  {author} {\bibinfo {author} {\bibfnamefont {S.}~\bibnamefont
  {Auer}}\ and\ \bibinfo {author} {\bibfnamefont {D.}~\bibnamefont {Frenkel}},\
  }\href@noop {} {\bibfield  {journal} {\bibinfo  {journal} {Nature}\ }\textbf
  {\bibinfo {volume} {409}},\ \bibinfo {pages} {1020} (\bibinfo {year}
  {2001})}\BibitemShut {NoStop}%
\bibitem [{\citenamefont {Punnathanam}\ and\ \citenamefont
  {Monson}(2006)}]{punnathanam2006crystal}%
  \BibitemOpen
  \bibfield  {author} {\bibinfo {author} {\bibfnamefont {S.}~\bibnamefont
  {Punnathanam}}\ and\ \bibinfo {author} {\bibfnamefont {P.}~\bibnamefont
  {Monson}},\ }\href@noop {} {\bibfield  {journal} {\bibinfo  {journal} {The
  Journal of chemical physics}\ }\textbf {\bibinfo {volume} {125}},\ \bibinfo
  {pages} {024508} (\bibinfo {year} {2006})}\BibitemShut {NoStop}%
\bibitem [{\citenamefont {Filion}\ \emph {et~al.}(2010)\citenamefont {Filion},
  \citenamefont {Hermes}, \citenamefont {Ni},\ and\ \citenamefont
  {Dijkstra}}]{filion2010crystal}%
  \BibitemOpen
  \bibfield  {author} {\bibinfo {author} {\bibfnamefont {L.}~\bibnamefont
  {Filion}}, \bibinfo {author} {\bibfnamefont {M.}~\bibnamefont {Hermes}},
  \bibinfo {author} {\bibfnamefont {R.}~\bibnamefont {Ni}}, \ and\ \bibinfo
  {author} {\bibfnamefont {M.}~\bibnamefont {Dijkstra}},\ }\href@noop {}
  {\bibfield  {journal} {\bibinfo  {journal} {The Journal of chemical physics}\
  }\textbf {\bibinfo {volume} {133}},\ \bibinfo {pages} {244115} (\bibinfo
  {year} {2010})}\BibitemShut {NoStop}%
\bibitem [{\citenamefont {Ni}\ \emph {et~al.}(2011)\citenamefont {Ni},
  \citenamefont {Smallenburg}, \citenamefont {Filion},\ and\ \citenamefont
  {Dijkstra}}]{ni2011crystal}%
  \BibitemOpen
  \bibfield  {author} {\bibinfo {author} {\bibfnamefont {R.}~\bibnamefont
  {Ni}}, \bibinfo {author} {\bibfnamefont {F.}~\bibnamefont {Smallenburg}},
  \bibinfo {author} {\bibfnamefont {L.}~\bibnamefont {Filion}}, \ and\ \bibinfo
  {author} {\bibfnamefont {M.}~\bibnamefont {Dijkstra}},\ }\href@noop {}
  {\bibfield  {journal} {\bibinfo  {journal} {Molecular Physics}\ }\textbf
  {\bibinfo {volume} {109}},\ \bibinfo {pages} {1213} (\bibinfo {year}
  {2011})}\BibitemShut {NoStop}%
\bibitem [{\citenamefont {Berryman}\ \emph {et~al.}(2016)\citenamefont
  {Berryman}, \citenamefont {Anwar}, \citenamefont {Dorosz},\ and\
  \citenamefont {Schilling}}]{berryman2016early}%
  \BibitemOpen
  \bibfield  {author} {\bibinfo {author} {\bibfnamefont {J.~T.}\ \bibnamefont
  {Berryman}}, \bibinfo {author} {\bibfnamefont {M.}~\bibnamefont {Anwar}},
  \bibinfo {author} {\bibfnamefont {S.}~\bibnamefont {Dorosz}}, \ and\ \bibinfo
  {author} {\bibfnamefont {T.}~\bibnamefont {Schilling}},\ }\href@noop {}
  {\bibfield  {journal} {\bibinfo  {journal} {The Journal of chemical physics}\
  }\textbf {\bibinfo {volume} {145}},\ \bibinfo {pages} {211901} (\bibinfo
  {year} {2016})}\BibitemShut {NoStop}%
\bibitem [{\citenamefont {Richard}\ and\ \citenamefont
  {Speck}(2018)}]{richard2018crystallization}%
  \BibitemOpen
  \bibfield  {author} {\bibinfo {author} {\bibfnamefont {D.}~\bibnamefont
  {Richard}}\ and\ \bibinfo {author} {\bibfnamefont {T.}~\bibnamefont
  {Speck}},\ }\href@noop {} {\bibfield  {journal} {\bibinfo  {journal} {The
  Journal of chemical physics}\ }\textbf {\bibinfo {volume} {148}},\ \bibinfo
  {pages} {124110} (\bibinfo {year} {2018})}\BibitemShut {NoStop}%
\bibitem [{\citenamefont {Wood}\ \emph {et~al.}(2018)\citenamefont {Wood},
  \citenamefont {Russo}, \citenamefont {Turci},\ and\ \citenamefont
  {Royall}}]{wood2018coupling}%
  \BibitemOpen
  \bibfield  {author} {\bibinfo {author} {\bibfnamefont {N.}~\bibnamefont
  {Wood}}, \bibinfo {author} {\bibfnamefont {J.}~\bibnamefont {Russo}},
  \bibinfo {author} {\bibfnamefont {F.}~\bibnamefont {Turci}}, \ and\ \bibinfo
  {author} {\bibfnamefont {C.~P.}\ \bibnamefont {Royall}},\ }\href@noop {}
  {\bibfield  {journal} {\bibinfo  {journal} {The Journal of chemical physics}\
  }\textbf {\bibinfo {volume} {149}},\ \bibinfo {pages} {204506} (\bibinfo
  {year} {2018})}\BibitemShut {NoStop}%
\bibitem [{\citenamefont {Espinosa}\ \emph {et~al.}(2019)\citenamefont
  {Espinosa}, \citenamefont {Vega}, \citenamefont {Valeriani}, \citenamefont
  {Frenkel},\ and\ \citenamefont {Sanz}}]{espinosa2019heterogeneous}%
  \BibitemOpen
  \bibfield  {author} {\bibinfo {author} {\bibfnamefont {J.~R.}\ \bibnamefont
  {Espinosa}}, \bibinfo {author} {\bibfnamefont {C.}~\bibnamefont {Vega}},
  \bibinfo {author} {\bibfnamefont {C.}~\bibnamefont {Valeriani}}, \bibinfo
  {author} {\bibfnamefont {D.}~\bibnamefont {Frenkel}}, \ and\ \bibinfo
  {author} {\bibfnamefont {E.}~\bibnamefont {Sanz}},\ }\href@noop {} {\bibfield
   {journal} {\bibinfo  {journal} {Soft Matter}\ }\textbf {\bibinfo {volume}
  {15}},\ \bibinfo {pages} {9625} (\bibinfo {year} {2019})}\BibitemShut
  {NoStop}%
\bibitem [{\citenamefont {Rintoul}\ and\ \citenamefont
  {Torquato}(1996)}]{rintoul1996computer}%
  \BibitemOpen
  \bibfield  {author} {\bibinfo {author} {\bibfnamefont {M.}~\bibnamefont
  {Rintoul}}\ and\ \bibinfo {author} {\bibfnamefont {S.}~\bibnamefont
  {Torquato}},\ }\href@noop {} {\bibfield  {journal} {\bibinfo  {journal} {The
  Journal of chemical physics}\ }\textbf {\bibinfo {volume} {105}},\ \bibinfo
  {pages} {9258} (\bibinfo {year} {1996})}\BibitemShut {NoStop}%
\bibitem [{\citenamefont {Speedy}(1998)}]{speedy1998hard}%
  \BibitemOpen
  \bibfield  {author} {\bibinfo {author} {\bibfnamefont {R.~J.}\ \bibnamefont
  {Speedy}},\ }\href@noop {} {\bibfield  {journal} {\bibinfo  {journal}
  {Molecular Physics}\ }\textbf {\bibinfo {volume} {95}},\ \bibinfo {pages}
  {169} (\bibinfo {year} {1998})}\BibitemShut {NoStop}%
\bibitem [{\citenamefont {Leocmach}\ and\ \citenamefont
  {Tanaka}(2012)}]{leocmach2012roles}%
  \BibitemOpen
  \bibfield  {author} {\bibinfo {author} {\bibfnamefont {M.}~\bibnamefont
  {Leocmach}}\ and\ \bibinfo {author} {\bibfnamefont {H.}~\bibnamefont
  {Tanaka}},\ }\href@noop {} {\bibfield  {journal} {\bibinfo  {journal} {Nature
  communications}\ }\textbf {\bibinfo {volume} {3}},\ \bibinfo {pages} {1}
  (\bibinfo {year} {2012})}\BibitemShut {NoStop}%
\bibitem [{\citenamefont {Sanz}\ \emph {et~al.}(2014)\citenamefont {Sanz},
  \citenamefont {Valeriani}, \citenamefont {Zaccarelli}, \citenamefont {Poon},
  \citenamefont {Cates},\ and\ \citenamefont {Pusey}}]{sanz2014avalanches}%
  \BibitemOpen
  \bibfield  {author} {\bibinfo {author} {\bibfnamefont {E.}~\bibnamefont
  {Sanz}}, \bibinfo {author} {\bibfnamefont {C.}~\bibnamefont {Valeriani}},
  \bibinfo {author} {\bibfnamefont {E.}~\bibnamefont {Zaccarelli}}, \bibinfo
  {author} {\bibfnamefont {W.~C.}\ \bibnamefont {Poon}}, \bibinfo {author}
  {\bibfnamefont {M.~E.}\ \bibnamefont {Cates}}, \ and\ \bibinfo {author}
  {\bibfnamefont {P.~N.}\ \bibnamefont {Pusey}},\ }\href@noop {} {\bibfield
  {journal} {\bibinfo  {journal} {Proceedings of the National Academy of
  Sciences}\ }\textbf {\bibinfo {volume} {111}},\ \bibinfo {pages} {75}
  (\bibinfo {year} {2014})}\BibitemShut {NoStop}%
\bibitem [{\citenamefont {Urbani}\ and\ \citenamefont
  {Zamponi}(2017)}]{urbani2017shear}%
  \BibitemOpen
  \bibfield  {author} {\bibinfo {author} {\bibfnamefont {P.}~\bibnamefont
  {Urbani}}\ and\ \bibinfo {author} {\bibfnamefont {F.}~\bibnamefont
  {Zamponi}},\ }\href@noop {} {\bibfield  {journal} {\bibinfo  {journal}
  {Physical review letters}\ }\textbf {\bibinfo {volume} {118}},\ \bibinfo
  {pages} {038001} (\bibinfo {year} {2017})}\BibitemShut {NoStop}%
\bibitem [{\citenamefont {Berthier}\ \emph {et~al.}(2017)\citenamefont
  {Berthier}, \citenamefont {Charbonneau}, \citenamefont {Coslovich},
  \citenamefont {Ninarello}, \citenamefont {Ozawa},\ and\ \citenamefont
  {Yaida}}]{berthier2017configurational}%
  \BibitemOpen
  \bibfield  {author} {\bibinfo {author} {\bibfnamefont {L.}~\bibnamefont
  {Berthier}}, \bibinfo {author} {\bibfnamefont {P.}~\bibnamefont
  {Charbonneau}}, \bibinfo {author} {\bibfnamefont {D.}~\bibnamefont
  {Coslovich}}, \bibinfo {author} {\bibfnamefont {A.}~\bibnamefont
  {Ninarello}}, \bibinfo {author} {\bibfnamefont {M.}~\bibnamefont {Ozawa}}, \
  and\ \bibinfo {author} {\bibfnamefont {S.}~\bibnamefont {Yaida}},\
  }\href@noop {} {\bibfield  {journal} {\bibinfo  {journal} {Proceedings of the
  National Academy of Sciences}\ }\textbf {\bibinfo {volume} {114}},\ \bibinfo
  {pages} {11356} (\bibinfo {year} {2017})}\BibitemShut {NoStop}%
\bibitem [{\citenamefont {L{\'a}zaro-L{\'a}zaro}\ \emph
  {et~al.}(2019)\citenamefont {L{\'a}zaro-L{\'a}zaro}, \citenamefont
  {Perera-Burgos}, \citenamefont {Laermann}, \citenamefont {Sentjabrskaja},
  \citenamefont {P{\'e}rez-{\'A}ngel}, \citenamefont {Laurati}, \citenamefont
  {Egelhaaf}, \citenamefont {Medina-Noyola}, \citenamefont {Voigtmann},
  \citenamefont {Casta{\~n}eda-Priego} \emph {et~al.}}]{lazaro2019glassy}%
  \BibitemOpen
  \bibfield  {author} {\bibinfo {author} {\bibfnamefont {E.}~\bibnamefont
  {L{\'a}zaro-L{\'a}zaro}}, \bibinfo {author} {\bibfnamefont {J.~A.}\
  \bibnamefont {Perera-Burgos}}, \bibinfo {author} {\bibfnamefont
  {P.}~\bibnamefont {Laermann}}, \bibinfo {author} {\bibfnamefont
  {T.}~\bibnamefont {Sentjabrskaja}}, \bibinfo {author} {\bibfnamefont
  {G.}~\bibnamefont {P{\'e}rez-{\'A}ngel}}, \bibinfo {author} {\bibfnamefont
  {M.}~\bibnamefont {Laurati}}, \bibinfo {author} {\bibfnamefont {S.~U.}\
  \bibnamefont {Egelhaaf}}, \bibinfo {author} {\bibfnamefont {M.}~\bibnamefont
  {Medina-Noyola}}, \bibinfo {author} {\bibfnamefont {T.}~\bibnamefont
  {Voigtmann}}, \bibinfo {author} {\bibfnamefont {R.}~\bibnamefont
  {Casta{\~n}eda-Priego}},  \emph {et~al.},\ }\href@noop {} {\bibfield
  {journal} {\bibinfo  {journal} {Physical Review E}\ }\textbf {\bibinfo
  {volume} {99}},\ \bibinfo {pages} {042603} (\bibinfo {year}
  {2019})}\BibitemShut {NoStop}%
\bibitem [{\citenamefont {Mar{\'\i}n-Aguilar}\ \emph
  {et~al.}(2020)\citenamefont {Mar{\'\i}n-Aguilar}, \citenamefont {Wensink},
  \citenamefont {Foffi},\ and\ \citenamefont
  {Smallenburg}}]{marin2020tetrahedrality}%
  \BibitemOpen
  \bibfield  {author} {\bibinfo {author} {\bibfnamefont {S.}~\bibnamefont
  {Mar{\'\i}n-Aguilar}}, \bibinfo {author} {\bibfnamefont {H.~H.}\ \bibnamefont
  {Wensink}}, \bibinfo {author} {\bibfnamefont {G.}~\bibnamefont {Foffi}}, \
  and\ \bibinfo {author} {\bibfnamefont {F.}~\bibnamefont {Smallenburg}},\
  }\href@noop {} {\bibfield  {journal} {\bibinfo  {journal} {Physical Review
  Letters}\ }\textbf {\bibinfo {volume} {124}},\ \bibinfo {pages} {208005}
  (\bibinfo {year} {2020})}\BibitemShut {NoStop}%
\bibitem [{\citenamefont {Boattini}\ \emph {et~al.}(2021)\citenamefont
  {Boattini}, \citenamefont {Smallenburg},\ and\ \citenamefont
  {Filion}}]{boattini2021averaging}%
  \BibitemOpen
  \bibfield  {author} {\bibinfo {author} {\bibfnamefont {E.}~\bibnamefont
  {Boattini}}, \bibinfo {author} {\bibfnamefont {F.}~\bibnamefont
  {Smallenburg}}, \ and\ \bibinfo {author} {\bibfnamefont {L.}~\bibnamefont
  {Filion}},\ }\href@noop {} {\bibfield  {journal} {\bibinfo  {journal} {Phys.
  Rev. Lett.}\ }\textbf {\bibinfo {volume} {127}},\ \bibinfo {pages} {088007}
  (\bibinfo {year} {2021})}\BibitemShut {NoStop}%
\bibitem [{\citenamefont {Boattini}\ \emph {et~al.}(2020)\citenamefont
  {Boattini}, \citenamefont {Mar{\'\i}n-Aguilar}, \citenamefont {Mitra},
  \citenamefont {Foffi}, \citenamefont {Smallenburg},\ and\ \citenamefont
  {Filion}}]{boattini2020autonomously}%
  \BibitemOpen
  \bibfield  {author} {\bibinfo {author} {\bibfnamefont {E.}~\bibnamefont
  {Boattini}}, \bibinfo {author} {\bibfnamefont {S.}~\bibnamefont
  {Mar{\'\i}n-Aguilar}}, \bibinfo {author} {\bibfnamefont {S.}~\bibnamefont
  {Mitra}}, \bibinfo {author} {\bibfnamefont {G.}~\bibnamefont {Foffi}},
  \bibinfo {author} {\bibfnamefont {F.}~\bibnamefont {Smallenburg}}, \ and\
  \bibinfo {author} {\bibfnamefont {L.}~\bibnamefont {Filion}},\ }\href@noop {}
  {\bibfield  {journal} {\bibinfo  {journal} {Nature communications}\ }\textbf
  {\bibinfo {volume} {11}},\ \bibinfo {pages} {1} (\bibinfo {year}
  {2020})}\BibitemShut {NoStop}%
\bibitem [{\citenamefont {Rosenbluth}\ and\ \citenamefont
  {Rosenbluth}(1954)}]{rosenbluth1954further}%
  \BibitemOpen
  \bibfield  {author} {\bibinfo {author} {\bibfnamefont {M.~N.}\ \bibnamefont
  {Rosenbluth}}\ and\ \bibinfo {author} {\bibfnamefont {A.~W.}\ \bibnamefont
  {Rosenbluth}},\ }\href@noop {} {\bibfield  {journal} {\bibinfo  {journal}
  {The Journal of Chemical Physics}\ }\textbf {\bibinfo {volume} {22}},\
  \bibinfo {pages} {881} (\bibinfo {year} {1954})}\BibitemShut {NoStop}%
\bibitem [{\citenamefont {Wood}\ and\ \citenamefont
  {Jacobson}(1957)}]{wood1957preliminary}%
  \BibitemOpen
  \bibfield  {author} {\bibinfo {author} {\bibfnamefont {W.~W.}\ \bibnamefont
  {Wood}}\ and\ \bibinfo {author} {\bibfnamefont {J.}~\bibnamefont
  {Jacobson}},\ }\href@noop {} {\bibfield  {journal} {\bibinfo  {journal} {The
  Journal of Chemical Physics}\ }\textbf {\bibinfo {volume} {27}},\ \bibinfo
  {pages} {1207} (\bibinfo {year} {1957})}\BibitemShut {NoStop}%
\bibitem [{\citenamefont {Frenkel}\ and\ \citenamefont
  {Smit}(2002)}]{bookfrenkel}%
  \BibitemOpen
  \bibfield  {author} {\bibinfo {author} {\bibfnamefont {D.}~\bibnamefont
  {Frenkel}}\ and\ \bibinfo {author} {\bibfnamefont {B.}~\bibnamefont {Smit}},\
  }\href@noop {} {\emph {\bibinfo {title} {Understanding Molecular Simulations:
  From Algorithms to Applications}}}\ (\bibinfo  {publisher} {Academic Press},\
  \bibinfo {address} {San Diego},\ \bibinfo {year} {2002})\BibitemShut
  {NoStop}%
\bibitem [{\citenamefont {Dress}\ and\ \citenamefont
  {Krauth}(1995)}]{dress1995cluster}%
  \BibitemOpen
  \bibfield  {author} {\bibinfo {author} {\bibfnamefont {C.}~\bibnamefont
  {Dress}}\ and\ \bibinfo {author} {\bibfnamefont {W.}~\bibnamefont {Krauth}},\
  }\href@noop {} {\bibfield  {journal} {\bibinfo  {journal} {Journal of Physics
  A: Mathematical and General}\ }\textbf {\bibinfo {volume} {28}},\ \bibinfo
  {pages} {L597} (\bibinfo {year} {1995})}\BibitemShut {NoStop}%
\bibitem [{\citenamefont {Almarza}(2009)}]{almarza2009cluster}%
  \BibitemOpen
  \bibfield  {author} {\bibinfo {author} {\bibfnamefont {N.~G.}\ \bibnamefont
  {Almarza}},\ }\href@noop {} {\bibfield  {journal} {\bibinfo  {journal} {The
  Journal of chemical physics}\ }\textbf {\bibinfo {volume} {130}},\ \bibinfo
  {pages} {184106} (\bibinfo {year} {2009})}\BibitemShut {NoStop}%
\bibitem [{\citenamefont {Ashton}\ \emph {et~al.}(2011)\citenamefont {Ashton},
  \citenamefont {Wilding}, \citenamefont {Roth},\ and\ \citenamefont
  {Evans}}]{ashton2011depletion}%
  \BibitemOpen
  \bibfield  {author} {\bibinfo {author} {\bibfnamefont {D.~J.}\ \bibnamefont
  {Ashton}}, \bibinfo {author} {\bibfnamefont {N.~B.}\ \bibnamefont {Wilding}},
  \bibinfo {author} {\bibfnamefont {R.}~\bibnamefont {Roth}}, \ and\ \bibinfo
  {author} {\bibfnamefont {R.}~\bibnamefont {Evans}},\ }\href@noop {}
  {\bibfield  {journal} {\bibinfo  {journal} {Physical Review E}\ }\textbf
  {\bibinfo {volume} {84}},\ \bibinfo {pages} {061136} (\bibinfo {year}
  {2011})}\BibitemShut {NoStop}%
\bibitem [{\citenamefont {Berthier}\ \emph {et~al.}(2016)\citenamefont
  {Berthier}, \citenamefont {Coslovich}, \citenamefont {Ninarello},\ and\
  \citenamefont {Ozawa}}]{berthier2016equilibrium}%
  \BibitemOpen
  \bibfield  {author} {\bibinfo {author} {\bibfnamefont {L.}~\bibnamefont
  {Berthier}}, \bibinfo {author} {\bibfnamefont {D.}~\bibnamefont {Coslovich}},
  \bibinfo {author} {\bibfnamefont {A.}~\bibnamefont {Ninarello}}, \ and\
  \bibinfo {author} {\bibfnamefont {M.}~\bibnamefont {Ozawa}},\ }\href@noop {}
  {\bibfield  {journal} {\bibinfo  {journal} {Physical review letters}\
  }\textbf {\bibinfo {volume} {116}},\ \bibinfo {pages} {238002} (\bibinfo
  {year} {2016})}\BibitemShut {NoStop}%
\bibitem [{\citenamefont {Bernard}\ \emph {et~al.}(2009)\citenamefont
  {Bernard}, \citenamefont {Krauth},\ and\ \citenamefont
  {Wilson}}]{bernard2009event}%
  \BibitemOpen
  \bibfield  {author} {\bibinfo {author} {\bibfnamefont {E.~P.}\ \bibnamefont
  {Bernard}}, \bibinfo {author} {\bibfnamefont {W.}~\bibnamefont {Krauth}}, \
  and\ \bibinfo {author} {\bibfnamefont {D.~B.}\ \bibnamefont {Wilson}},\
  }\href@noop {} {\bibfield  {journal} {\bibinfo  {journal} {Physical Review
  E}\ }\textbf {\bibinfo {volume} {80}},\ \bibinfo {pages} {056704} (\bibinfo
  {year} {2009})}\BibitemShut {NoStop}%
\bibitem [{\citenamefont {Klement}\ and\ \citenamefont
  {Engel}(2019)}]{klement2019efficient}%
  \BibitemOpen
  \bibfield  {author} {\bibinfo {author} {\bibfnamefont {M.}~\bibnamefont
  {Klement}}\ and\ \bibinfo {author} {\bibfnamefont {M.}~\bibnamefont
  {Engel}},\ }\href@noop {} {\bibfield  {journal} {\bibinfo  {journal} {The
  Journal of chemical physics}\ }\textbf {\bibinfo {volume} {150}},\ \bibinfo
  {pages} {174108} (\bibinfo {year} {2019})}\BibitemShut {NoStop}%
\bibitem [{\citenamefont {Sanz}\ and\ \citenamefont
  {Marenduzzo}(2010)}]{sanz2010dynamic}%
  \BibitemOpen
  \bibfield  {author} {\bibinfo {author} {\bibfnamefont {E.}~\bibnamefont
  {Sanz}}\ and\ \bibinfo {author} {\bibfnamefont {D.}~\bibnamefont
  {Marenduzzo}},\ }\href@noop {} {\bibfield  {journal} {\bibinfo  {journal}
  {The Journal of chemical physics}\ }\textbf {\bibinfo {volume} {132}},\
  \bibinfo {pages} {194102} (\bibinfo {year} {2010})}\BibitemShut {NoStop}%
\bibitem [{\citenamefont {Cuetos}\ and\ \citenamefont
  {Patti}(2015)}]{cuetos2015equivalence}%
  \BibitemOpen
  \bibfield  {author} {\bibinfo {author} {\bibfnamefont {A.}~\bibnamefont
  {Cuetos}}\ and\ \bibinfo {author} {\bibfnamefont {A.}~\bibnamefont {Patti}},\
  }\href@noop {} {\bibfield  {journal} {\bibinfo  {journal} {Physical Review
  E}\ }\textbf {\bibinfo {volume} {92}},\ \bibinfo {pages} {022302} (\bibinfo
  {year} {2015})}\BibitemShut {NoStop}%
\bibitem [{\citenamefont {Alder}\ and\ \citenamefont
  {Wainwright}(1957)}]{alder1957phase}%
  \BibitemOpen
  \bibfield  {author} {\bibinfo {author} {\bibfnamefont {B.~J.}\ \bibnamefont
  {Alder}}\ and\ \bibinfo {author} {\bibfnamefont {T.~E.}\ \bibnamefont
  {Wainwright}},\ }\href@noop {} {\bibfield  {journal} {\bibinfo  {journal}
  {The Journal of chemical physics}\ }\textbf {\bibinfo {volume} {27}},\
  \bibinfo {pages} {1208} (\bibinfo {year} {1957})}\BibitemShut {NoStop}%
\bibitem [{\citenamefont {Scala}\ \emph {et~al.}(2007)\citenamefont {Scala},
  \citenamefont {Voigtmann},\ and\ \citenamefont
  {De~Michele}}]{scala2007event}%
  \BibitemOpen
  \bibfield  {author} {\bibinfo {author} {\bibfnamefont {A.}~\bibnamefont
  {Scala}}, \bibinfo {author} {\bibfnamefont {T.}~\bibnamefont {Voigtmann}}, \
  and\ \bibinfo {author} {\bibfnamefont {C.}~\bibnamefont {De~Michele}},\
  }\href@noop {} {\bibfield  {journal} {\bibinfo  {journal} {The Journal of
  chemical physics}\ }\textbf {\bibinfo {volume} {126}},\ \bibinfo {pages}
  {134109} (\bibinfo {year} {2007})}\BibitemShut {NoStop}%
\bibitem [{\citenamefont {Scala}(2012)}]{scala2012event}%
  \BibitemOpen
  \bibfield  {author} {\bibinfo {author} {\bibfnamefont {A.}~\bibnamefont
  {Scala}},\ }\href@noop {} {\bibfield  {journal} {\bibinfo  {journal}
  {Physical Review E}\ }\textbf {\bibinfo {volume} {86}},\ \bibinfo {pages}
  {026709} (\bibinfo {year} {2012})}\BibitemShut {NoStop}%
\bibitem [{\citenamefont {Paul}(2007)}]{paul2007complexity}%
  \BibitemOpen
  \bibfield  {author} {\bibinfo {author} {\bibfnamefont {G.}~\bibnamefont
  {Paul}},\ }\href@noop {} {\bibfield  {journal} {\bibinfo  {journal} {Journal
  of Computational Physics}\ }\textbf {\bibinfo {volume} {221}},\ \bibinfo
  {pages} {615} (\bibinfo {year} {2007})}\BibitemShut {NoStop}%
\bibitem [{\citenamefont {Donev}\ \emph
  {et~al.}(2005{\natexlab{a}})\citenamefont {Donev}, \citenamefont {Torquato},\
  and\ \citenamefont {Stillinger}}]{donev2005neighbor}%
  \BibitemOpen
  \bibfield  {author} {\bibinfo {author} {\bibfnamefont {A.}~\bibnamefont
  {Donev}}, \bibinfo {author} {\bibfnamefont {S.}~\bibnamefont {Torquato}}, \
  and\ \bibinfo {author} {\bibfnamefont {F.~H.}\ \bibnamefont {Stillinger}},\
  }\href@noop {} {\bibfield  {journal} {\bibinfo  {journal} {Journal of
  computational physics}\ }\textbf {\bibinfo {volume} {202}},\ \bibinfo {pages}
  {737} (\bibinfo {year} {2005}{\natexlab{a}})}\BibitemShut {NoStop}%
\bibitem [{\citenamefont {Lubachevsky}(1991)}]{lubachevsky1991simulate}%
  \BibitemOpen
  \bibfield  {author} {\bibinfo {author} {\bibfnamefont {B.~D.}\ \bibnamefont
  {Lubachevsky}},\ }\href@noop {} {\bibfield  {journal} {\bibinfo  {journal}
  {Journal of Computational Physics}\ }\textbf {\bibinfo {volume} {94}},\
  \bibinfo {pages} {255} (\bibinfo {year} {1991})}\BibitemShut {NoStop}%
\bibitem [{\citenamefont {Bannerman}\ \emph {et~al.}(2011)\citenamefont
  {Bannerman}, \citenamefont {Sargant},\ and\ \citenamefont
  {Lue}}]{bannerman2011dynamo}%
  \BibitemOpen
  \bibfield  {author} {\bibinfo {author} {\bibfnamefont {M.~N.}\ \bibnamefont
  {Bannerman}}, \bibinfo {author} {\bibfnamefont {R.}~\bibnamefont {Sargant}},
  \ and\ \bibinfo {author} {\bibfnamefont {L.}~\bibnamefont {Lue}},\
  }\href@noop {} {\bibfield  {journal} {\bibinfo  {journal} {Journal of
  computational chemistry}\ }\textbf {\bibinfo {volume} {32}},\ \bibinfo
  {pages} {3329} (\bibinfo {year} {2011})}\BibitemShut {NoStop}%
\bibitem [{\citenamefont {Rapaport}\ and\ \citenamefont
  {Rapaport}(2004)}]{rapaport2004art}%
  \BibitemOpen
  \bibfield  {author} {\bibinfo {author} {\bibfnamefont {D.~C.}\ \bibnamefont
  {Rapaport}}\ and\ \bibinfo {author} {\bibfnamefont {D.~C.~R.}\ \bibnamefont
  {Rapaport}},\ }\href@noop {} {\emph {\bibinfo {title} {The art of molecular
  dynamics simulation}}}\ (\bibinfo  {publisher} {Cambridge university press},\
  \bibinfo {year} {2004})\BibitemShut {NoStop}%
\bibitem [{\citenamefont {Mar{\i}n}\ \emph {et~al.}(1993)\citenamefont
  {Mar{\i}n}, \citenamefont {Risso},\ and\ \citenamefont
  {Cordero}}]{marin1993efficient}%
  \BibitemOpen
  \bibfield  {author} {\bibinfo {author} {\bibfnamefont {M.}~\bibnamefont
  {Mar{\i}n}}, \bibinfo {author} {\bibfnamefont {D.}~\bibnamefont {Risso}}, \
  and\ \bibinfo {author} {\bibfnamefont {P.}~\bibnamefont {Cordero}},\
  }\href@noop {} {\bibfield  {journal} {\bibinfo  {journal} {Journal of
  Computational Physics}\ }\textbf {\bibinfo {volume} {109}},\ \bibinfo {pages}
  {306} (\bibinfo {year} {1993})}\BibitemShut {NoStop}%
\bibitem [{\citenamefont {Mar{\'\i}n}\ and\ \citenamefont
  {Cordero}(1995)}]{marin1995empirical}%
  \BibitemOpen
  \bibfield  {author} {\bibinfo {author} {\bibfnamefont {M.}~\bibnamefont
  {Mar{\'\i}n}}\ and\ \bibinfo {author} {\bibfnamefont {P.}~\bibnamefont
  {Cordero}},\ }\href@noop {} {\bibfield  {journal} {\bibinfo  {journal}
  {Computer Physics Communications}\ }\textbf {\bibinfo {volume} {92}},\
  \bibinfo {pages} {214} (\bibinfo {year} {1995})}\BibitemShut {NoStop}%
\bibitem [{\citenamefont {Khan}\ and\ \citenamefont
  {Herbordt}(2011)}]{khan2011parallel}%
  \BibitemOpen
  \bibfield  {author} {\bibinfo {author} {\bibfnamefont {M.~A.}\ \bibnamefont
  {Khan}}\ and\ \bibinfo {author} {\bibfnamefont {M.~C.}\ \bibnamefont
  {Herbordt}},\ }\href@noop {} {\bibfield  {journal} {\bibinfo  {journal}
  {Journal of computational physics}\ }\textbf {\bibinfo {volume} {230}},\
  \bibinfo {pages} {6563} (\bibinfo {year} {2011})}\BibitemShut {NoStop}%
\bibitem [{NEC()}]{NECgithub}%
  \BibitemOpen
  \href@noop {} {\enquote {\bibinfo {title}
  {https://github.com/glotzerlab/hoomd-blue/tree/newtonian\_event\_chain\_monte\_carlo},}\
  }\bibinfo {note} {[Accessed: 14-10-2021]}\BibitemShut {NoStop}%
\bibitem [{\citenamefont {Anderson}\ \emph {et~al.}(2020)\citenamefont
  {Anderson}, \citenamefont {Glaser},\ and\ \citenamefont
  {Glotzer}}]{anderson2020hoomd}%
  \BibitemOpen
  \bibfield  {author} {\bibinfo {author} {\bibfnamefont {J.~A.}\ \bibnamefont
  {Anderson}}, \bibinfo {author} {\bibfnamefont {J.}~\bibnamefont {Glaser}}, \
  and\ \bibinfo {author} {\bibfnamefont {S.~C.}\ \bibnamefont {Glotzer}},\
  }\href@noop {} {\bibfield  {journal} {\bibinfo  {journal} {Computational
  Materials Science}\ }\textbf {\bibinfo {volume} {173}},\ \bibinfo {pages}
  {109363} (\bibinfo {year} {2020})}\BibitemShut {NoStop}%
\bibitem [{\citenamefont {Klement}\ \emph {et~al.}(2021)\citenamefont
  {Klement}, \citenamefont {Lee}, \citenamefont {Anderson},\ and\ \citenamefont
  {Engel}}]{klement2021newtonian}%
  \BibitemOpen
  \bibfield  {author} {\bibinfo {author} {\bibfnamefont {M.}~\bibnamefont
  {Klement}}, \bibinfo {author} {\bibfnamefont {S.}~\bibnamefont {Lee}},
  \bibinfo {author} {\bibfnamefont {J.~A.}\ \bibnamefont {Anderson}}, \ and\
  \bibinfo {author} {\bibfnamefont {M.}~\bibnamefont {Engel}},\ }\href@noop {}
  {\bibfield  {journal} {\bibinfo  {journal} {arXiv preprint arXiv:2104.06829}\
  } (\bibinfo {year} {2021})}\BibitemShut {NoStop}%
\bibitem [{\citenamefont {Anderson}\ \emph {et~al.}(2013)\citenamefont
  {Anderson}, \citenamefont {Jankowski}, \citenamefont {Grubb}, \citenamefont
  {Engel},\ and\ \citenamefont {Glotzer}}]{anderson2013massively}%
  \BibitemOpen
  \bibfield  {author} {\bibinfo {author} {\bibfnamefont {J.~A.}\ \bibnamefont
  {Anderson}}, \bibinfo {author} {\bibfnamefont {E.}~\bibnamefont {Jankowski}},
  \bibinfo {author} {\bibfnamefont {T.~L.}\ \bibnamefont {Grubb}}, \bibinfo
  {author} {\bibfnamefont {M.}~\bibnamefont {Engel}}, \ and\ \bibinfo {author}
  {\bibfnamefont {S.~C.}\ \bibnamefont {Glotzer}},\ }\href@noop {} {\bibfield
  {journal} {\bibinfo  {journal} {Journal of Computational Physics}\ }\textbf
  {\bibinfo {volume} {254}},\ \bibinfo {pages} {27} (\bibinfo {year}
  {2013})}\BibitemShut {NoStop}%
\bibitem [{\citenamefont {Engel}\ \emph {et~al.}(2013)\citenamefont {Engel},
  \citenamefont {Anderson}, \citenamefont {Glotzer}, \citenamefont {Isobe},
  \citenamefont {Bernard},\ and\ \citenamefont {Krauth}}]{engel2013hard}%
  \BibitemOpen
  \bibfield  {author} {\bibinfo {author} {\bibfnamefont {M.}~\bibnamefont
  {Engel}}, \bibinfo {author} {\bibfnamefont {J.~A.}\ \bibnamefont {Anderson}},
  \bibinfo {author} {\bibfnamefont {S.~C.}\ \bibnamefont {Glotzer}}, \bibinfo
  {author} {\bibfnamefont {M.}~\bibnamefont {Isobe}}, \bibinfo {author}
  {\bibfnamefont {E.~P.}\ \bibnamefont {Bernard}}, \ and\ \bibinfo {author}
  {\bibfnamefont {W.}~\bibnamefont {Krauth}},\ }\href@noop {} {\bibfield
  {journal} {\bibinfo  {journal} {Physical Review E}\ }\textbf {\bibinfo
  {volume} {87}},\ \bibinfo {pages} {042134} (\bibinfo {year}
  {2013})}\BibitemShut {NoStop}%
\bibitem [{\citenamefont {Miller}\ and\ \citenamefont
  {Luding}(2004)}]{miller2004event}%
  \BibitemOpen
  \bibfield  {author} {\bibinfo {author} {\bibfnamefont {S.}~\bibnamefont
  {Miller}}\ and\ \bibinfo {author} {\bibfnamefont {S.}~\bibnamefont
  {Luding}},\ }\href@noop {} {\bibfield  {journal} {\bibinfo  {journal}
  {Journal of Computational Physics}\ }\textbf {\bibinfo {volume} {193}},\
  \bibinfo {pages} {306} (\bibinfo {year} {2004})}\BibitemShut {NoStop}%
\bibitem [{\citenamefont {Donev}\ \emph
  {et~al.}(2005{\natexlab{b}})\citenamefont {Donev}, \citenamefont {Torquato},\
  and\ \citenamefont {Stillinger}}]{donev2005neighbor2}%
  \BibitemOpen
  \bibfield  {author} {\bibinfo {author} {\bibfnamefont {A.}~\bibnamefont
  {Donev}}, \bibinfo {author} {\bibfnamefont {S.}~\bibnamefont {Torquato}}, \
  and\ \bibinfo {author} {\bibfnamefont {F.~H.}\ \bibnamefont {Stillinger}},\
  }\href@noop {} {\bibfield  {journal} {\bibinfo  {journal} {Journal of
  computational physics}\ }\textbf {\bibinfo {volume} {202}},\ \bibinfo {pages}
  {765} (\bibinfo {year} {2005}{\natexlab{b}})}\BibitemShut {NoStop}%
\bibitem [{\citenamefont {Hern{\'a}ndez de~la Pe{\~n}a}\ \emph
  {et~al.}(2007)\citenamefont {Hern{\'a}ndez de~la Pe{\~n}a}, \citenamefont
  {van Zon}, \citenamefont {Schofield},\ and\ \citenamefont
  {Opps}}]{hernandez2007discontinuous}%
  \BibitemOpen
  \bibfield  {author} {\bibinfo {author} {\bibfnamefont {L.}~\bibnamefont
  {Hern{\'a}ndez de~la Pe{\~n}a}}, \bibinfo {author} {\bibfnamefont
  {R.}~\bibnamefont {van Zon}}, \bibinfo {author} {\bibfnamefont
  {J.}~\bibnamefont {Schofield}}, \ and\ \bibinfo {author} {\bibfnamefont
  {S.~B.}\ \bibnamefont {Opps}},\ }\href@noop {} {\bibfield  {journal}
  {\bibinfo  {journal} {The Journal of chemical physics}\ }\textbf {\bibinfo
  {volume} {126}},\ \bibinfo {pages} {074105} (\bibinfo {year}
  {2007})}\BibitemShut {NoStop}%
\bibitem [{\citenamefont {Smallenburg}\ \emph {et~al.}(2012)\citenamefont
  {Smallenburg}, \citenamefont {Filion}, \citenamefont {Marechal},\ and\
  \citenamefont {Dijkstra}}]{smallenburg2012vacancy}%
  \BibitemOpen
  \bibfield  {author} {\bibinfo {author} {\bibfnamefont {F.}~\bibnamefont
  {Smallenburg}}, \bibinfo {author} {\bibfnamefont {L.}~\bibnamefont {Filion}},
  \bibinfo {author} {\bibfnamefont {M.}~\bibnamefont {Marechal}}, \ and\
  \bibinfo {author} {\bibfnamefont {M.}~\bibnamefont {Dijkstra}},\ }\href@noop
  {} {\bibfield  {journal} {\bibinfo  {journal} {Proceedings of the National
  Academy of Sciences}\ }\textbf {\bibinfo {volume} {109}},\ \bibinfo {pages}
  {17886} (\bibinfo {year} {2012})}\BibitemShut {NoStop}%
\bibitem [{\citenamefont {Smallenburg}\ and\ \citenamefont
  {Sciortino}(2013)}]{smallenburg2013liquids}%
  \BibitemOpen
  \bibfield  {author} {\bibinfo {author} {\bibfnamefont {F.}~\bibnamefont
  {Smallenburg}}\ and\ \bibinfo {author} {\bibfnamefont {F.}~\bibnamefont
  {Sciortino}},\ }\href@noop {} {\bibfield  {journal} {\bibinfo  {journal}
  {Nature Physics}\ }\textbf {\bibinfo {volume} {9}},\ \bibinfo {pages} {554}
  (\bibinfo {year} {2013})}\BibitemShut {NoStop}%
\bibitem [{\citenamefont {Mar{\'\i}n-Aguilar}\ \emph
  {et~al.}(2019)\citenamefont {Mar{\'\i}n-Aguilar}, \citenamefont {Wensink},
  \citenamefont {Foffi},\ and\ \citenamefont {Smallenburg}}]{marin2019slowing}%
  \BibitemOpen
  \bibfield  {author} {\bibinfo {author} {\bibfnamefont {S.}~\bibnamefont
  {Mar{\'\i}n-Aguilar}}, \bibinfo {author} {\bibfnamefont {H.~H.}\ \bibnamefont
  {Wensink}}, \bibinfo {author} {\bibfnamefont {G.}~\bibnamefont {Foffi}}, \
  and\ \bibinfo {author} {\bibfnamefont {F.}~\bibnamefont {Smallenburg}},\
  }\href@noop {} {\bibfield  {journal} {\bibinfo  {journal} {Soft matter}\
  }\textbf {\bibinfo {volume} {15}},\ \bibinfo {pages} {9886} (\bibinfo {year}
  {2019})}\BibitemShut {NoStop}%
\end{thebibliography}

\end{document}